\documentclass[%
 aip,
 amsmath,amssymb,
 reprint,%
]{revtex4-1}

\usepackage{graphicx}
\usepackage{dcolumn}
\usepackage{bm}

\usepackage[utf8]{inputenc}
\usepackage[T1]{fontenc}
\usepackage{mathptmx}
\usepackage{etoolbox}
\usepackage{multirow}

\usepackage{xcolor, color}
\usepackage{tabularx}
\usepackage{comment}
\usepackage{soul}

\makeatletter
\def\@email#1#2{%
 \endgroup
 \patchcmd{\titleblock@produce}
  {\frontmatter@RRAPformat}
  {\frontmatter@RRAPformat{\produce@RRAP{*#1\href{mailto:#2}{#2}}}\frontmatter@RRAPformat}
  {}{}
}%
\makeatother
\begin{document}

\preprint{JAP/Sentaurus-FKT}

\title[TCAD Modeling of GaN HEMTs]{A comparison of a commercial hydrodynamics TCAD solver and Fermi kinetics transport convergence for GaN HEMTs}
\author{Ashwin Tunga}
\email{tunga2@illinois.edu}
\affiliation{Holonyak Micro and Nanotechnology Laboratory, University of Illinois at Urbana-Champaign, Urbana, IL, 61801, USA}
\author{Kexin Li}%
\affiliation{Holonyak Micro and Nanotechnology Laboratory, University of Illinois at Urbana-Champaign, Urbana, IL, 61801, USA}
\author{Nicholas C. Miller}%
\affiliation{Air Force Research Laboratory Sensors Directorate, 2241 Avionics Cir., Wright-Patterson AFB, OH, 45433, USA}%
\author{Matt Grupen}%
\affiliation{Air Force Research Laboratory Sensors Directorate, 2241 Avionics Cir., Wright-Patterson AFB, OH, 45433, USA}%
\author{John D. Albrecht}%
\affiliation{Department of Electrical and Computer Engineering, Michigan State University, 428 S. Shaw Lane, East Lansing, MI, 48824, USA}%
\author{Shaloo Rakheja}
\affiliation{Holonyak Micro and Nanotechnology Laboratory, University of Illinois at Urbana-Champaign, Urbana, IL, 61801, USA}
\email{rakheja@illinois.edu}

\date{\today}

\begin{abstract}
Various simulations of a GaN HEMT are used to study the behaviors of two different energy-transport models: the Fermi kinetics transport model and a hydrodynamics transport model as it is implemented in the device simulator Sentaurus from Synopsys. The electron transport and heat flow equations of the respective solvers are described in detail. The differences in the description of electron flux and the discretization methods are highlighted. Next, the transport models are applied to the same simulated device structure using identical meshes, boundary conditions, and material parameters. Static simulations show the numerical convergence of Fermi kinetics to be consistently quadratic or faster, whereas the hydrodynamic model is often sub-quadratic. Further comparisons of large signal transient simulations reveal the hydrodynamic model produces certain anomalous electron ensemble behaviors within the transistor structure. The fundamentally different electron dynamics produced by the two models suggest an underlying cause for their different numerical convergence characteristics. 
\end{abstract}

\maketitle

\section{\label{sec:level1} Introduction}

Recent decades have seen increased interest in GaN devices for power and radio frequency (RF) electronics.\cite{hooteoEmergingGaNTechnologies2021, baligaGalliumNitrideDevices2013} Numerical simulation is a necessary step for the design and analysis of these devices. Robust simulation models rely on producing predictive results. Accurate simulation of these devices must capture the important physical processes involved and the carrier transport mechanisms. While it is possible to obtain reasonable results using ad hoc solutions despite incorrectly modeling the physics, these results are not predictive and do not aid in device design.~\cite{stettlerCriticalExaminationAssumptions1993, budeMOSFETModelingBallistic2000, grasserInvestigationSpuriousVelocity2001}  Furthermore, these numerical methods may be problematic for large-signal transient simulations, which are particularly important for RF power applications. This paper seeks to explore some numerical behaviors of established charge transport solvers for RF GaN HEMTs.

Charge transport in a semiconductor is described by the Boltzmann transport equation (BTE). The BTE is an integro-differential equation that describes the evolution of the non-equilibrium distribution function of electrons under the influence of an external electric field. The BTE is given as \cite{KHess}
\begin{equation}
\begin{split}
&\frac{\partial f(\bm{k}, \bm{r}, t)}{\partial t}+\bm{v} \cdot \nabla_{r} f(\bm{k}, \bm{r}, t)+\frac{\bm{F}}{\hbar} \cdot \nabla_{k} f(\bm{k}, \bm{r}, t) \\
&=\left(\frac{\partial f(\bm{k}, \bm{r}, t)}{\partial t}\right)_{\text {coll }},
\end{split}
\end{equation}
where $f$ is the distribution function, $\bm{F}$ is the external force, $\bm{v}$ is the group velocity, $\bm{k}$ is the wavevector in the reciprocal space, $\bm{r}$ is position vector in the real space, $t$ is the time, and the term on the right-hand side represents the rate of change of $f$ due to the collisions between the carriers.

Monte Carlo solvers are a useful statistical method to solve the BTE. The Monte Carlo technique generally employed is the ensemble Monte Carlo (EMC) method.~\cite{fischettiMonteCarloAnalysis1988} The accurate results produced by this method come at the cost of immense computational load. To reduce the computational load encountered by the EMC, the hybrid cellular automaton/Monte Carlo (CA/MC) algorithm was developed. The CA/MC scheme attempts to speed-up the simulation by pre-calculating the scattering rates and storing them in a table. It was shown that the CA/MC method required about 0.2-0.5 CPU-sec per iteration, up to 25 times faster than the traditional EMC.~\cite{saranitiHybridFullbandCellular2000} However, this speed improvement offered by the hybrid scheme is still not comparable to the shorter simulation time taken by deterministic solvers.

Deterministic solvers are an alternative approach that does not rely on random sampling to solve the BTE. While the deterministic methods are generally much faster than the Monte Carlo methods, they rely on physical approximations to the BTE
to obtain the solutions.~\cite{selberherrAnalysisSimulationSemiconductor1984} Deterministic solvers convert the BTE into an equivalent infinite series of differential equations in phase space by applying the method of moments over the reciprocal space. Truncating the series after a finite number of moments typically results in one solution variable more than the available equations. Therefore, a closure relation is required to make the system of equations tractable.~\cite{grasserReviewHydrodynamicEnergytransport2003}

Scharfetter and Gummel~\cite{scharfetterLargesignalAnalysisSilicon1969} developed one of the first deterministic Boltzmann solvers. They proposed a robust discretization of the drift-diffusion (DD) model with physical approximations. However, as the device dimensions were scaled, the DD model failed to account for the large electric fields and the hot-carrier effects that arise, e.g., in FET channels. These effects are typically accounted for in the DD model by empirically modifying the mobility.~\cite{penzinKineticVelocityModel2017}

The DD model was extended by Stratton \cite{strattonDiffusionHotCold1962} and Blotekjaer \cite{blotekjaerTransportEquationsElectrons1970} to include a balance equation for the average energy of the carriers. These models are generally referred to as energy transport models, and the hydrodynamic model is a prime example.~\cite{grasserReviewHydrodynamicEnergytransport2003} Hydrodynamic models typically use the parabolic band approximation to describe the kinetic energy vs momentum dispersion relation. This approximation fails to hold when the electric fields in the device reach large magnitudes. To account for this, the model relies on empirical field-dependent and temperature-dependent mobility models. The commonly used closure relation in hydrodynamic models is the description of heat flow using Fourier’s law and an approximate thermal conductivity.

Synopsys Sentaurus is an industry-leading suite of semiconductor device and process simulation tools. Sentaurus has been widely used to simulate devices for various applications including power electronics,~\cite{straussTCADMethodologySimulation2014,krausPhysicsBasedCompactModel2016} RF electronics,~\cite{ajayanInvestigationDCRFBreakdown2018} and CMOS technology.~\cite{zhengFinFETEvolutionStackedNanowire2015,wu3DTCADSimulation2017} Sentaurus provides a wide range of simulation packages to handle the vast application space and support a streamlined simulation workflow. The range of packages available includes Sentaurus Process for process simulation, Sentaurus Device Monte Carlo, Sentaurus Device Electromagnetic Wave Solver, and, the package under study in this work, Sentaurus Device (SDevice).~\cite{SentaurusDeviceUser2020}
The history of SDevice can be traced back to the development of DESSIS in 1992. DESSIS was developed by merging HFIELDS from the University of Bologna,~\cite{baccarani1985hfields} SIMUL from ETH Zurich,~\cite{litsios1993mixed} BONSIM from Bosch, and in collaboration with ST Microelectronics.~\cite{baccaraniDeviceSimulationSmart1996} In 1993, Integrated Systems Engineering (ISE) was founded and assumed responsibility for the simulator. ISE merged with Synopsys in 2004, and DESSIS became the basis of SDevice.~\cite{sarkarTechnologyComputerAided2018}
SDevice supports several carrier transport models that can be used depending on the device under investigation. The hydrodynamic model, implemented in SDevice, will be investigated in this work.

FKT is also an energy transport based BTE solver that employs the method of moments but uses an alternative treatment of the electronic heat flow. It substitutes the idea of thermal conductivity with the heat capacity of an ideal Fermi gas to be used as the closure relation.~\cite{grupenAlternativeTreatmentHeat2009} It has been demonstrated that FKT, unlike the hydrodynamic model, could incorporate the electronic band structure~\cite{grupenEnergyTransportModel2011} and full-wave EM.~\cite{grupenThreeDimensionalFullWaveElectromagnetics2014} FKT has also demonstrated large-signal RF simulations~\cite{miller_large-signal_2016} and defect dynamics simulations\cite{grupen_reproducing_2019}. More details on the solver are reported in the literature.~\cite{miller_delaunay-voronoi_2016, miller2017large, miller_computational_2018} 

This paper examines the physical applicability of both the FKT and the hydrodynamic model employed by Sentaurus, hereafter referred to as commercial hydrodynamic transport (CHT) solver. First, the system of equations and the discretization scheme employed by the solvers are presented in Section~\ref{sec:comp_bg}. Section~\ref{sec:simresults} presents the GaN HEMT structure and the simulation parameters used in the respective simulators. Next, the thermal equilibrium simulations are performed and the results are compared. The output characteristics and the electron temperatures of the HEMT structure, produced by the non-equilibrium simulations, are compared.  Further, the differences in the simulation frameworks are investigated by analyzing the numerical convergence behavior after varying the simulation setup, to ensure similar simulation conditions for the purpose of comparison. Finally, the effect of varying the polarization sheet charge density at the AlGaN/GaN interface on the convergence characteristics and the transient simulation behavior of the simulators are investigated.

\section{Computational Background} \label{sec:comp_bg}



The semiconductor device simulators considered here are both of the
energy transport variety. As such, they have a great deal in common.
For example, they both tessellate a problem domain with a suitable mesh
comprised of triangles for 2D simulations or tetrahedra for 3D
simulations.
This mesh conforms to the simulated device's geometry and is refined
appropriately in regions where solution variables change rapidly, such
as material interfaces and metallurgical junctions.
Solution variables assigned to each mesh point include an
electrostatic potential to account for voltage drops and the
corresponding electric fields throughout the device. To each mesh
point is also assigned a Fermi-Dirac distribution for an electron gas
occupying the semiconductor's conduction band to account for currents
flowing through the device under the influence of the internal
fields. Integrals of the distribution function over the density of
conduction band states can provide the particle and energy densities
of mobile electrons throughout the device. These densities can be
treated as additional solution variables at each mesh point, or they
can be equivalently associated with alternative solution variables
such as the Fermi distribution's chemical potential (Fermi level) and
temperature.

Both device simulators calculate the solution variables by solving
Poisson's equation, the electron continuity equation, and the
conservation of electron energy at each mesh point using the
box-integration method \cite{selberherrAnalysisSimulationSemiconductor1984}. The electrostatic potentials are approximated
as piecewise linear along the edges connecting neighboring mesh
points, and the electron fluxes are approximated as piecewise
constant between mesh points. The electron particle and kinetic energy
fluxes can be derived from the 1st and 3rd moments of the BTE,
respectively \cite{KHess}. Both simulators represent these fluxes in a discrete form
on the device mesh using the powerful Scharfetter-Gummel method \cite{selberherrAnalysisSimulationSemiconductor1984,scharfetterLargesignalAnalysisSilicon1969}.
Because the mobile electrons are treated as ideal Fermi
gases, both simulators also account for the electronic heat that flows
between neighboring electron distributions.

Although the simulators use these common methodologies, there are some
differences in how they are implemented that lead to significant
differences in the simulation results and numerical convergence
characteristics. Principal differences include the precise forms of
the electron fluxes obtained from moments of the BTE, the way
Scharfetter-Gummel discretization is applied to these fluxes, and the
way electronic heat flow is treated. The following details these
differences along with the associated simulation results and numerical
behaviors. For simplicity, effective mass and constant mobility approximations will be used.

\subsection{CHT Simulator}

For the sake of simplicity, we will limit ourselves to the consideration of only electrons in steady-state conditions.
Additionally, the lattice temperature, $T_L$ is set to 300 K to disentangle the effects of electron heating and lattice heating, resolve convergence issues, and enable a fair comparison of CHT against FKT.  
The electron current density, modeled using CHT, is given as~\cite{SentaurusDeviceUser2020}
\begin{equation}
    \vec{J}_n = \mu_n \left[n \nabla E_C + kT_n \nabla n - nkT_n \nabla \ln \gamma_n + \lambda_n f_n^\mathrm{td} k n \nabla T_n\right].
    \label{eq:Jn}
\end{equation}
The first term in the above equation accounts for the spatial variation of electrostatic potential, while the remainder terms account for the gradient of concentration and the carrier temperature gradient. 
$\mu_n$ is the electron mobility, $n$ is the electron concentration, and $T_n$ denotes the temperature of electrons. 
The thermal diffusion constant $f_n^\mathrm{td}$ is set equal to zero, which corresponds to the Stratton model.~\cite{stratton1962diffusion}  
The parameters $\gamma_n$ and $\lambda_n$ account for the Fermi-Dirac statistics and are given as
\begin{align}
        \gamma_n = \frac{n}{N_C} \exp (-\eta_n),\\
        \lambda_n = \frac{\mathcal{F}_{1/2}(\eta_n)}{\mathcal{F}_{-1/2}(\eta_n)}.
        \label{eqn:lambda}
\end{align}
$\mathcal{F}_{j}(x)$ is the Blakemore Fermi-Dirac integral of order-$j$. The argument $\eta_n$ is the reduced Fermi energy and is defined as $\eta_n = (F_n-E_C)/kT_n$. The effective conduction band density of states in the semiconductor is specified by $N_C$. 
CHT provides an option, in lieu of $T_n$, to use a blended temperature in (\ref{eq:Jn}) according to $T_n \rightarrow gT_n + (1-g)T_L$, where $T_L$ is the lattice temperature and $g$ is a dimensionless parameter between zero and unity. For all 
CHT simulations reported in this paper, $g$ is set equal to unity. 

The energy balance equation is given as 
\begin{equation}
    \frac{\partial W_n}{\partial t} + \vec{\nabla} \cdot \vec{S}_n = \vec{J}_n \cdot \vec{\nabla}\frac{E_C}{q} + \frac{dW_n}{dt}|_\mathrm{coll.},
    \label{eq:ebal}
\end{equation}
where $W_n$ is the electron energy density, and $\vec{S}_n$ is the energy flux. 
$W_n$ comprises a thermal and a convective term; however, CHT implements $W_n$ without the convective term. That is, $W_n = n w_n = n \left(3kT_n/2\right)$. 
The term $\vec{J}_n \cdot \vec{\nabla}\frac{E_C}{q}$ represents the energy delivered to the carriers by the field per unit time, while $\frac{dW_n}{dt}|_\mathrm{coll.}$ is the net energy gained from the lattice via phonon interactions.   
$\vec{S}_n$ is given as 
\begin{equation}
 \vec{S}_n = - \frac{5 r_n \lambda_n}{2}\left[\frac{kT_n}{q} \vec{J}_n + f_n^\mathrm{hf} \hat{\kappa}_n \nabla T_n\right], \label{eqn:sentSn}
\end{equation}
and 
\begin{equation}
    \hat{\kappa}_n = \frac{k^2}{q} n \mu_n T_n.
\end{equation}
Here, we use the default parameter values for $r_n$ and $f_n^\mathrm{hf}$ as listed in Table~\ref{tab:sent_param}. 
These parameters allow us to change the relative contributions of the convective and diffusive terms in the energy flux. 
With the default parameters, the prefactor for the diffusive term becomes 
\begin{equation}
    \kappa_n = \frac{3}{2} \frac{k^2 \lambda_n}{q} n \mu_n T_n.
\end{equation}

The collision term in the energy balance is given as 
\begin{equation}
    \frac{dW_n}{dt}_\mathrm{coll.} = - \zeta_n \frac{W_n-W_{n0}}{\tau_{e}}.
    \label{eqn:sentCol}
\end{equation}
$W_{n0}$ is the equilibrium electron energy density, and $\tau_{e}$ is the \emph{energy relaxation time} for electrons. The parameter $\zeta_n$ is adjusted to improve the stability of the numerical algorithm and is given as 
\begin{equation}
    \zeta_n = 1+ \frac{n_\mathrm{min}}{n} \left[\frac{n_\mathrm{0}}{n_\mathrm{min}}\right]^{\mathrm{max}[0, (T_L-T_n)/100\text{ K}]},
\end{equation}
with $n_\mathrm{min}$ and $n_0$ are adjustable density parameters to aid convergence.  
Since $T_n \geq T_L$ for our device, $n_0$ does not impact $\zeta_n$. The default value of $n_\mathrm{min}$ is set to $10^3$ cm$^{-3}$, which leads to $\zeta_n \approx 1$ for all test cases reported here.  
While $\tau_e$ is energy-dependent and would lead to a modification of the diffusive energy flux, here we treat it as an energy-independent parameter. However, CHT provides empirical methods, such as spline approximation of energy relaxation time over energy, to account for $\tau_{e}$'s energy dependence. In this study, we vary $\tau_e$ in the fs to ps range to analyze its impact on I-V curves and electron heating. Additionally, we elucidate that $\tau_{e}$'s impact on I-V convergence in CHT. Detailed results are discussed in Sec.~\ref{sec:simresults}.


To discretize the electron fluxes, CHT uses the ingenious
Scharfettel-Gummel method
\cite{selberherrAnalysisSimulationSemiconductor1984,scharfetterLargesignalAnalysisSilicon1969}.
In its original form, this method assumed
the electric field $\mathcal{E} = (E_{C,1}-E_{C,0})/(qL)$ and the electron flux $J_n$
were constant on a mesh edge of length $L$ connecting mesh points 0
and 1. With these assumptions, the phenomenological drift-diffusion
equation was treated as a first-order differential equation for the
electron density $n$. Solving this differential equation for
variations in $n$ along the mesh edge then produced a discrete
form for the electron flux, given in (\ref{eq:Jn_sen_dis}), that proved to be numerically robust.
\begin{equation}
J_n = \frac{\mu_n}{\ell_{ij}} kT_n \left(n_i B(u_i-u_j)-n_j B(u_j-u_i)\right),
    \label{eq:Jn_sen_dis}
\end{equation}
where $i$ and $j$ are the major mesh points, $\mu_n$ is the electron mobility (assumed constant between the mesh points), $\ell_{ij}$ is the length of the mesh edge, $B(x) = x/(e^x-1)$ is the Bernoulli function, and $u_i-u_j = q\mathcal{E}\ell_{ij}/kT_n$, and $\mathcal{E}$ is constant between the mesh points. For high electron densities, CHT may also add some additional terms to (\ref{eq:Jn_sen_dis}) to account for Fermi-Dirac degeneracy.

%
%

\begin{table}[]
\caption{\label{tab:sentaurus_pars} Parameters in CHT.}
    \begin{ruledtabular}
    \begin{tabular}{c c p{135 pt}}
        \textbf{Parameter} & \textbf{Default Value} & \textbf{Meaning} \\
        \hline
        $r_n$ & 0.6 & Parameter in energy flux\\
        $f_n^{hf}$ & 1 & Parameter in energy flux\\
        $g$ & 1 & Parameter to select carrier temperature $(T_n \rightarrow gT_n + (1-g T_L))$\\
        $f_n^{td}$ & 0 & Thermal diffusion coefficient\\
        $n_\mathrm{min}$ & $10^3$ cm$^{-3}$& Parameter to improve numeric stability\\
    \end{tabular}
    \end{ruledtabular}
    \label{tab:sent_param}
\end{table}

\subsection{Fermi Kinetics Transport}

Because FKT, like CHT, assumes an energy-dependent electron
distribution function at points in the simulation domain, electron
fluxes are computed from moments of the BTE\@. This
method of moments is a powerful technique for inferring information
about the part of the distribution function odd in momentum-space,
such as net particle flux, from the even part of the distribution
function. For example, drift-diffusion flux, which is often presented
phenomenologically, can be derived from the first moment of the BTE
assuming a Maxwell-Boltzmann distribution function, constant electron
temperature, and constant mobility \cite{KHess}. In general, FKT uses
electron fluxes derived for a Fermi-Dirac distribution, non-constant
temperature, non-parabolic bands, and an energy-dependent mobility
computed from electron scattering rates \cite{grupenEnergyTransportModel2011}.
To simplify comparisons with CHT, the electron
current density from the first
moment assuming parabolic bands and constant mobility is given by
\begin{eqnarray}
\vec{J}_n &=& \mu_n\frac{(2m_n)^{3/2}}{3\pi^2\hbar^3}\left[
(kT_n)^{3/2}F_{3/2}'(\eta_n)\nabla E_C +
\right. \nonumber \\
&& \mbox{\hspace{0.75in}} \left.
\nabla (kT_n)^{5/2}F_{3/2}(\eta_n)\right], \label{eqn:fktJn}
\end{eqnarray}
where $F_j(\eta_n)$ is the integral over energy $E$ of the
electron Fermi function times $E^j$,
and $F_j'(\eta_n)$ is its derivative with
respect to $\eta_n$.
This integral is proportional to the Fermi integrals in
(\ref{eqn:lambda})
$\mathcal{F}_{j}(\eta_n) = F_j(\eta_n)/\Gamma(j+1)$.
The electron internal/kinetic energy flux is
likewise derived from the third moment of the BTE,
\begin{eqnarray}
\vec{S}_n^{\rm kin} &=& -\frac{\mu_n}{q}\frac{(2m_n)^{3/2}}{3\pi^2\hbar^3}
\left[(kT_n)^{5/2}F_{5/2}'(\eta_n)\nabla E_C +
\right. \nonumber \\
&& \mbox{\hspace{0.875in}} \left.
\nabla (kT_n)^{7/2}F_{7/2}(\eta_n)\right]. \label{eqn:fktSn}
\end{eqnarray}

Like CHT, FKT uses the Scharfetter-Gummel discretization,
but it treats (\ref{eqn:fktJn}) as a
first order differential equation for the quantity
$(kT_n)^{5/2}F_{3/2}$. Solving this differential equation produces a
discretized electron current density on the mesh edge given by
\begin{eqnarray}
J_n &=& -\frac{\mu_n}{L}
\frac{(2m_n)^{3/2}}{3\pi^2\hbar^3}\left[
B(\xi_n)(kT_0)^{5/2}F_{3/2}(\eta_{n,0}) - \right. \nonumber \\
&& \left. \mbox{\hspace{0.9375in}}
B(-\xi_n)(kT_1)^{5/2}F_{3/2}(\eta_{n,1})\right] \nonumber \\
&=& -q\left(J_{0\rightarrow1} - J_{1\rightarrow0}\right)
\label{eqn:fktSGJ}
\end{eqnarray}
\begin{equation}
\xi_n = \frac{\Delta(E_C)}{\Delta(F_n-E_C)}
\left[\ln F_{3/2}(\eta_{n,1}) - \ln F_{3/2}(\eta_{n,0})\right],
\end{equation}
where the Bernoulli function $B(\xi_n) = \xi_n/[\exp(\xi_n)-1]$,
and $\Delta(x) \equiv x_1 - x_0$.
The same procedure is applied to (\ref{eqn:fktSn}) to obtain the
Scharfetter-Gummel discretization of the electron kinetic energy flux
\begin{eqnarray}
S_n^{\rm kin} &=& \frac{\mu_n}{qL}
\frac{(2m_n)^{3/2}}{3\pi^2\hbar^3}\left[
B(\xi_E)(kT_0)^{7/2}F_{5/2}(\eta_{n,0}) - \right. \nonumber \\
&& \left. \mbox{\hspace{0.875in}}
B(-\xi_E)(kT_1)^{7/2}F_{5/2}(\eta_{n,1})\right] \nonumber \\
&=& S_{0\rightarrow1}^{\rm kin} - S_{1\rightarrow0}^{\rm kin}
\label{eqn:fktSGS}
\end{eqnarray}
\begin{equation}
\xi_E = \frac{\Delta(E_C)}{\Delta(F_n-E_C)}
\left[\ln F_{5/2}(\eta_{n,1}) - \ln F_{5/2}(\eta_{n,0})\right].
\end{equation}

Although FKT particle and kinetic energy fluxes as well as the flux
discretization schemes differ somewhat from the corresponding
quantities in CHT, perhaps the most significant departure is the
treatment of electronic heat flow. It is evident from
(\ref{eqn:sentSn}) that CHT treats electronic heat flow with
Fourier's law using an electronic thermal conductivity related to the
electrical conductivity.
FKT, on the other hand, exploits the kinetics of ideal Fermi gases \cite{grupenAlternativeTreatmentHeat2009}.
Because electron ensembles at points inside the simulation are assumed
to have an energy-dependent distribution, i.e., independent of the
direction of degenerate electron momentum vectors $\bm{k}$, the electron
ensembles are symmetric in $\bm{k}$-space. This is the defining feature
of an ideal gas. When simulated electron fluxes move along a mesh edge
from an initial mesh point with Fermi level $(F_n-E_C)_0$ and
temperature $T_{n,0}$ to a final mesh point with Fermi level
$(F_n-E_C)_1$ and temperature $T_{n,1}$, the electrons must thermalize
in order to change their temperature from $T_{n,0}$ to $T_{n,1}$. This
thermalization process is achieved through electron-electron
scattering, and it is the source of electronic heat flow between the
electron ensembles/gases. The amount of heat $H$ needed to change the
temperature of an ideal Fermi gas from initial temperature $T_i$ to
final temperature $T_f$ is precisely defined by the
thermodynamic identity \cite{CKittel}
\begin{equation}
H = W_n(T_f) - W_n(T_i) - (F_n-E_C)[n(T_f) - n(T_i)].
\end{equation}
Determining the net electronic heat flux on the mesh edge requires the
rate electrons change from $T_{n,0}$ to $T_{n,1}$ and the rate electrons
change from $T_{n,1}$ to $T_{n,0}$. These rates are simply
$J_{0\rightarrow1}$ and $J_{1\rightarrow0}$ from (\ref{eqn:fktSGJ}),
respectively.
Therefore, the electronic heat flux flowing from point 0 to
point 1 is given by
\begin{eqnarray}
S_{0\rightarrow1}^{\rm heat} &=& S_{0\rightarrow1}^{\rm kin}(T_{n,0}) -
S_{0\rightarrow1}^{\rm kin}(T_{n,1}) -
\nonumber \\ &&
(F_n-E_C)_0\left[
J_{0\rightarrow1}(T_{n,0}) -
J_{0\rightarrow1}(T_{n,1})\right],
\end{eqnarray}
the electronic heat flux flowing from point 1 to
point 0 is given by
\begin{eqnarray}
S_{1\rightarrow0}^{\rm heat} &=& S_{1\rightarrow0}^{\rm kin}(T_{n,1}) -
S_{1\rightarrow0}^{\rm kin}(T_{n,0}) -
\nonumber \\ &&
(F_n-E_C)_1\left[
J_{1\rightarrow0}(T_{n,1}) -
J_{1\rightarrow0}(T_{n,0})\right],
\end{eqnarray}
and the net electronic heat flux along the edge is
$S_n^{\rm heat} = S_{0\rightarrow1}^{\rm heat} - S_{1\rightarrow0}^{\rm heat}$.
Combining this heat flow with the kinetic energy flux gives the total
electron energy flux along the edge connecting mesh points 0 and 1,
$S_n = S_n^{\rm kin} + S_n^{\rm heat}$.

FKT uses the same Joule heat CHT uses in
(\ref{eq:ebal}). FKT typically computes the collision operator that
determines the rate hot electrons transfer energy to the semiconductor
lattice by integrating the phonon scattering rates over electron
energy. However, for the sake of the comparisons with CHT
presented here, FKT was modified to use (\ref{eqn:sentCol}).

\section{Comparisons of Simulation Results}
\label{sec:simresults}

\begin{figure}[h!]
    \centering
    \includegraphics[width=3.5in]{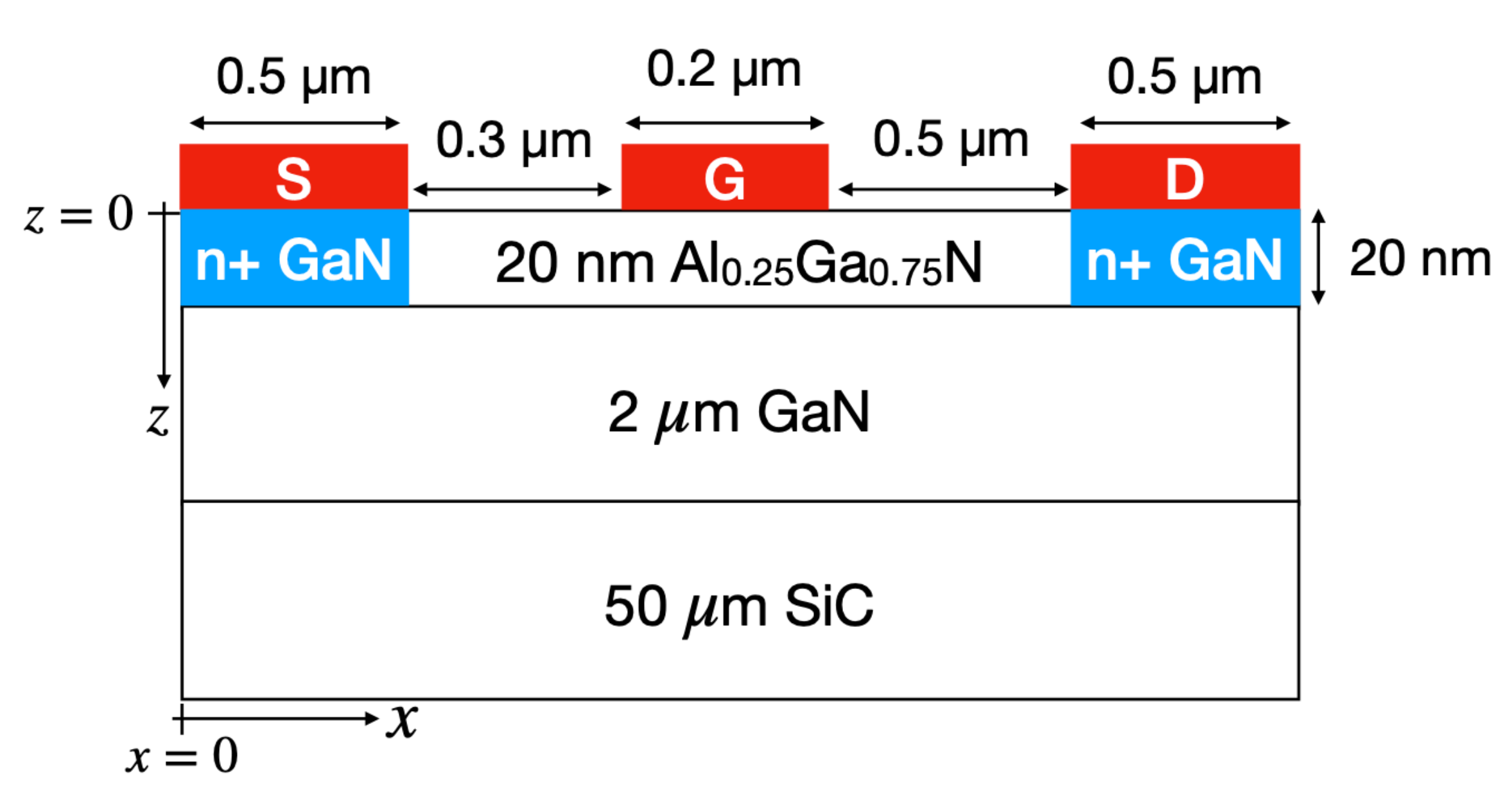}
    \caption{Cross-section of a GaN device simulated in FKT and CHT. The 2DEG exists in GaN at the interface between AlGaN (barrier) and GaN (channel). Figure is not drawn to scale.}
    \label{fig:gan_cross}
\end{figure}
\begin{figure}
    \centering
    \includegraphics[width=0.7\linewidth]{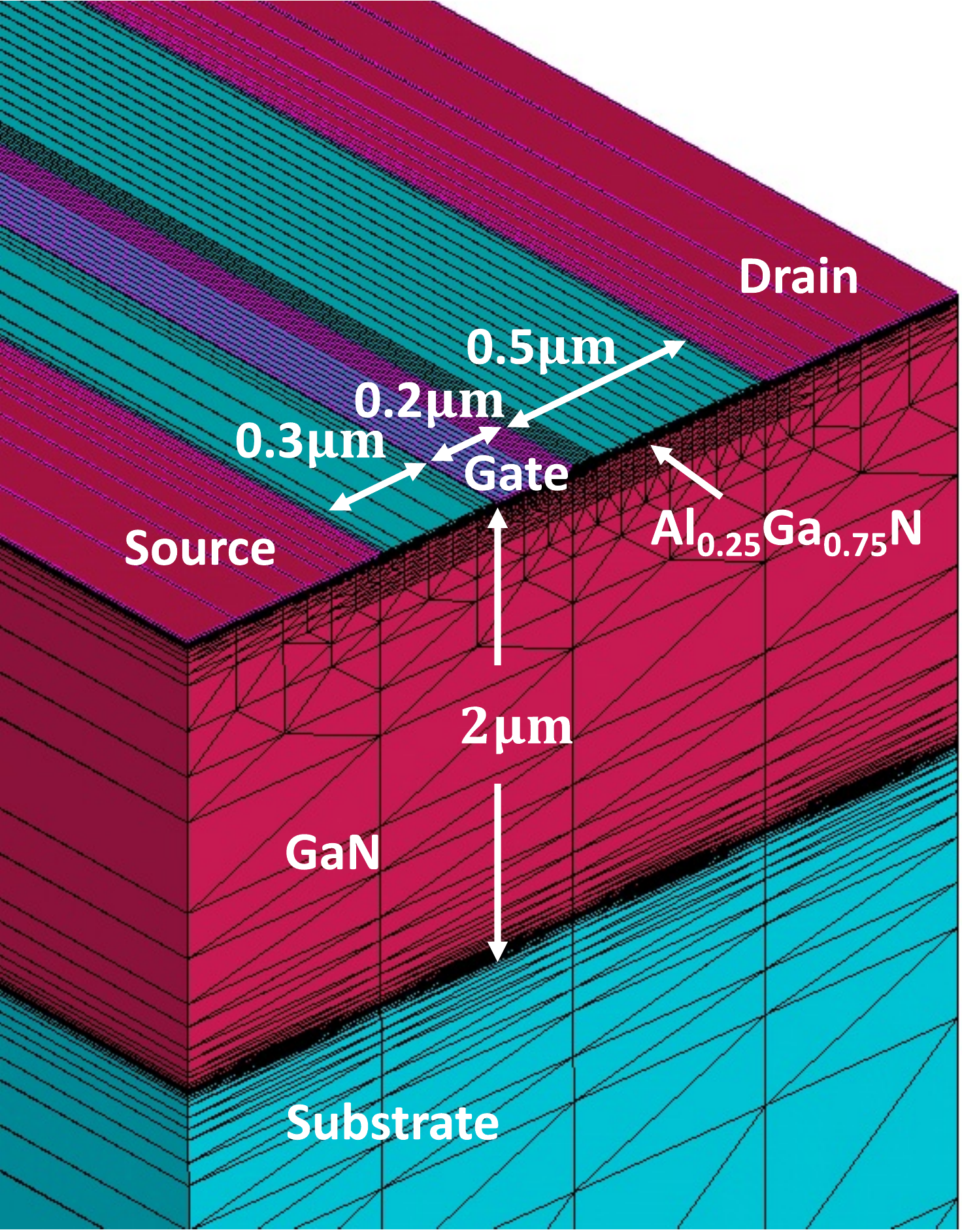}
    \caption{Mesh structure of the GaN HEMT device, shown in Fig. \ref{fig:gan_cross}, used for the simulations in both CHT and FKT. The mesh is refined near the 2DEG in the channel, and near the material interfaces, to obtain accurate results.}
    \label{fig:mesh}
\end{figure}

The GaN HEMT structure simulated in FKT and CHT is illustrated in Fig.~\ref{fig:gan_cross}. 
The n$^+$ GaN source/drain contacts are doped with $10^{19}$ cm$^{-3}$ donors and complete dopant ionization is assumed. 
The GaN region is semi-insulating containing $9\times 10^{15}$ cm$^{-3}$ fully ionized acceptor atoms compensated with $10^{16}$ cm$^{-3}$ donor-like traps located at an energy level 0.6 eV below $E_C$ and capture cross-section of $10^{-19}$ m$^2$.
The Schottky barrier height at the gate contact boundary condition is assumed to be 1.5 eV, while the source and drain contacts are ohmic boundary conditions.
For both simulators, AlGaN and SiC are treated as insulators, which means that carrier transport equations are not solved in these regions. All the simulations for both the simulators were run on the same mesh, shown in Fig. \ref{fig:mesh}.

\addtolength{\tabcolsep}{1pt}
\begin{table*}
\caption{\label{tab:material_pars} Material parameters that are common in both CHT and FKT.}
\begin{tabularx}{\textwidth}{c X X X p{90 pt} X X}
\hline \hline
Material & Dielectric constant & Electron Affinity \newline (eV) & Bandgap \newline (eV) & Electron mobility \newline $(\mathrm{cm}^2/\mathrm{V\,s})$ & Effective mass \newline ($\mathrm{m}_0$)\footnote{$\mathrm{m}_0$ is the free electron mass}\\
\hline
AlGaN & 8.9 & 3.55 & 4.14 & --- & ---\\
GaN & 8.9 & 4.1 & 3.457 & 440 & 0.2\\
SiC & 9.66 & --- & --- & --- & ---\\
\hline \hline
\end{tabularx}
\end{table*}
\addtolength{\tabcolsep}{-1pt}


\subsection{Thermal Equilibrium}
\begin{figure}[h!]
    \centering
    \includegraphics[width=0.42\textwidth]{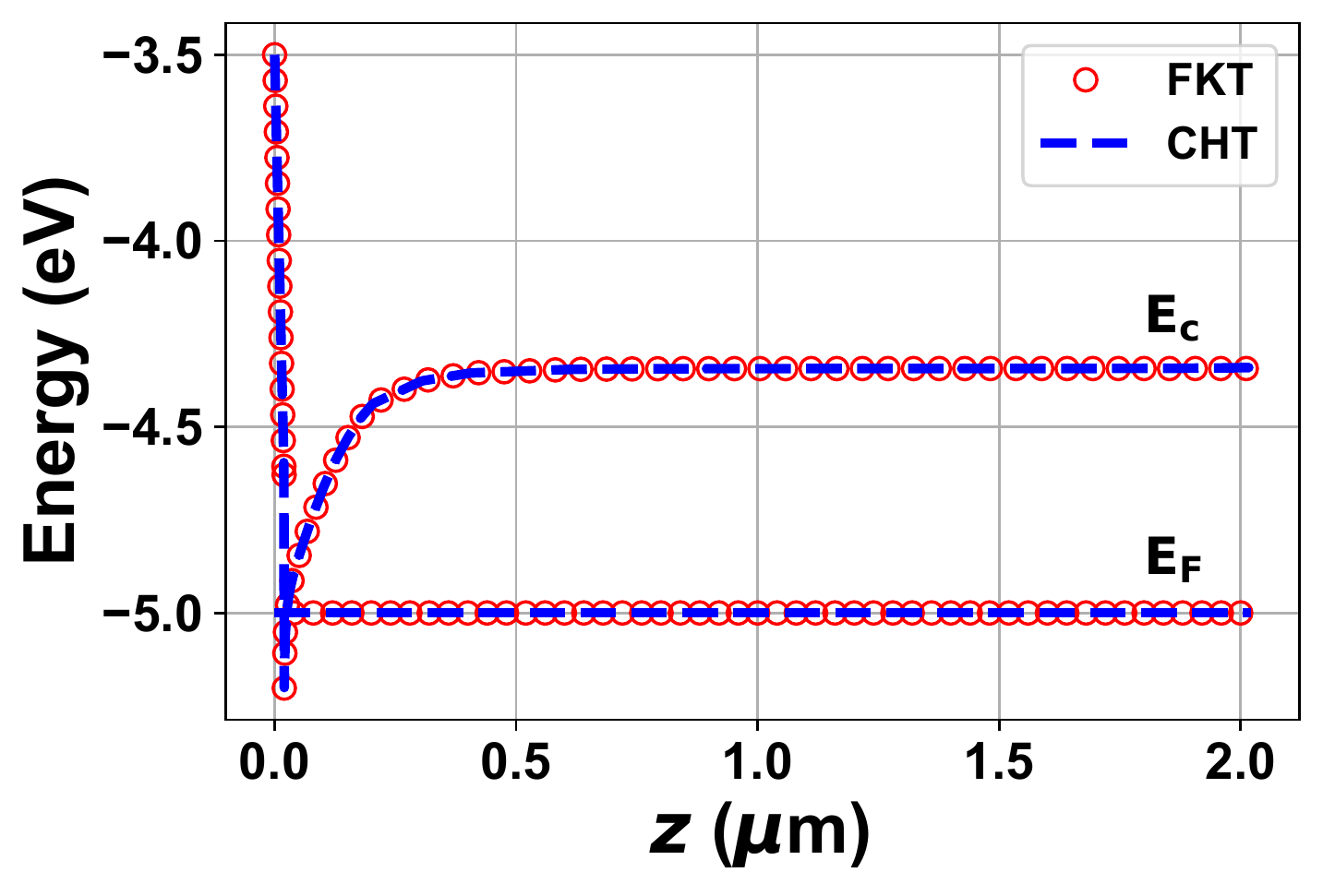}
    \caption{Comparison of the equilibrium band diagram produced by CHT and FKT along the z-direction (vertical) at $\mathrm{x} = 0.9$~$\mu \textrm{m}$. $\mathrm{E_C}$ denotes the conduction band edge and $\mathrm{E_F}$ refers to the Fermi level at thermal equilibrium.}
    \label{fig:eqbm_bnd_z}
\end{figure}

\begin{figure}[h!]
    \centering
    \includegraphics[width=0.42\textwidth]{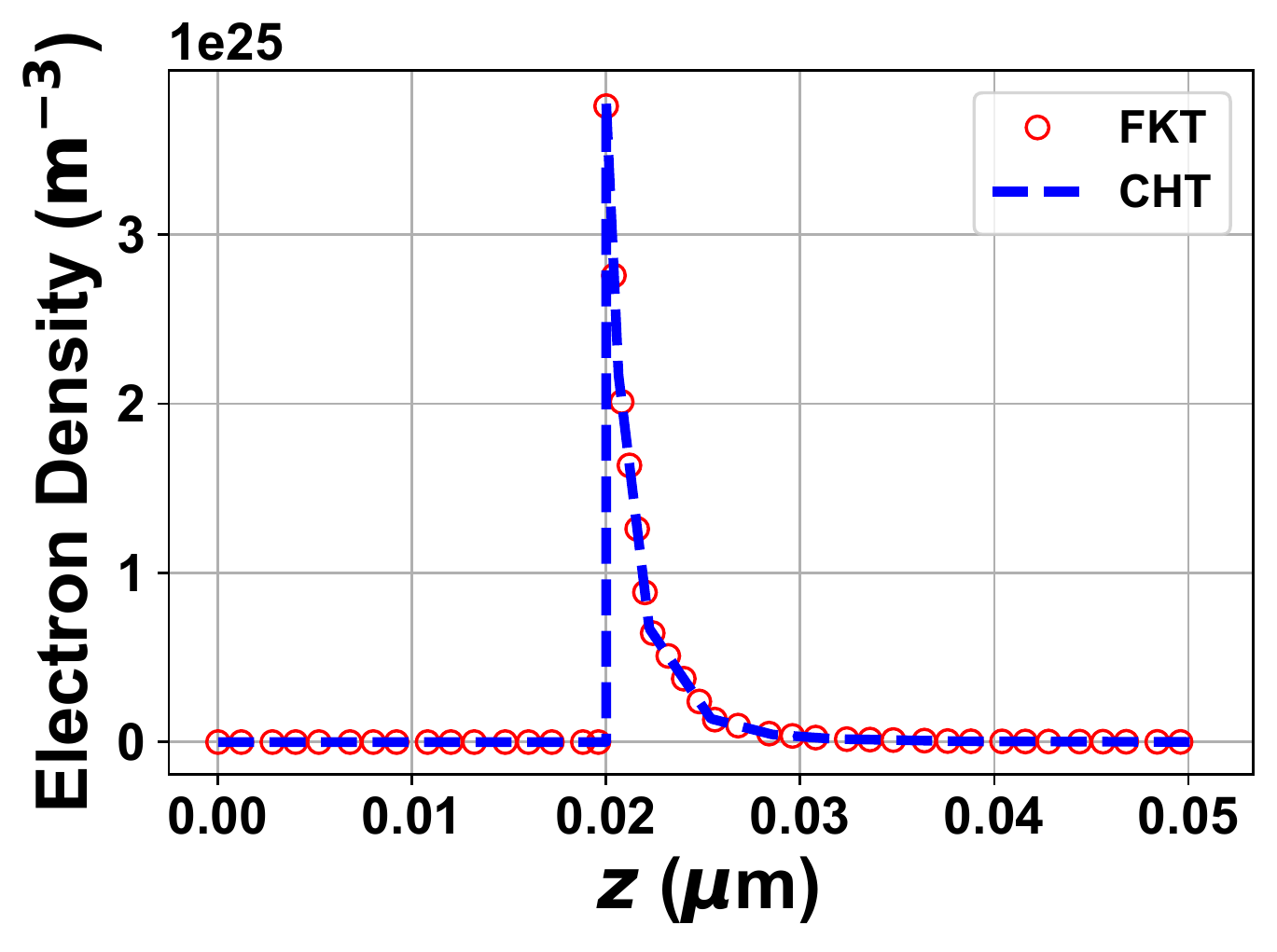}
    \caption{Comparison of the equilibrium electron density produced by CHT and FKT along the z-direction (vertical) at $\mathrm{x} = 0.9$~$\mu \textrm{m}$.}
    \label{fig:eqbm_eDens_z}
\end{figure}

\begin{figure}[h!]
    \centering
    \includegraphics[width=0.42\textwidth]{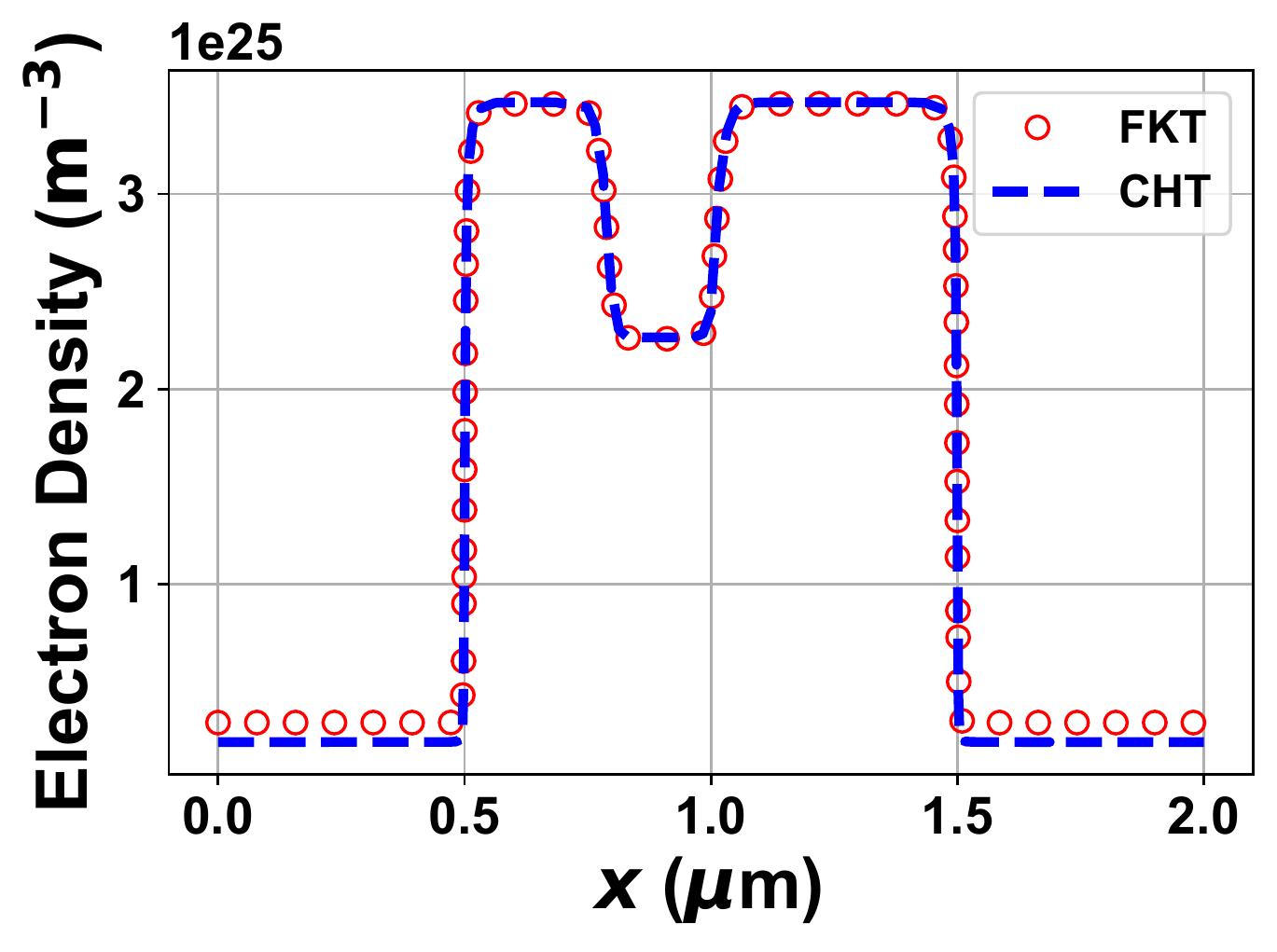}
    \caption{Comparison of the equilibrium electron density produced by CHT and FKT along the x-direction (channel) at $6$~nm below the AlGaN/GaN interface.}
    \label{fig:eqbm_eDens_x}
\end{figure}

The thermal equilibrium simulation 
does not depend on the transport model used. Therefore, one would expect identical equilibrium results from the two simulators, given the mesh, the material and simulation parameters for the GaN HEMT are kept alike. Keeping the aforementioned conditions identical, we calibrate the two simulators and verify that they produce similar results. Table \ref{tab:material_pars} shows the material parameters that are kept uniform. Figures~\ref{fig:eqbm_bnd_z}, \ref{fig:eqbm_eDens_z}, \ref{fig:eqbm_eDens_x} show the z-direction band diagram, z-direction electron density and the x-direction electron density plots, respectively. As expected, excellent agreement between the simulators is obtained. 

\subsection{Non-Equilibrium Simulations}



\begin{figure}
    \centering
    \includegraphics[width=0.42\textwidth]{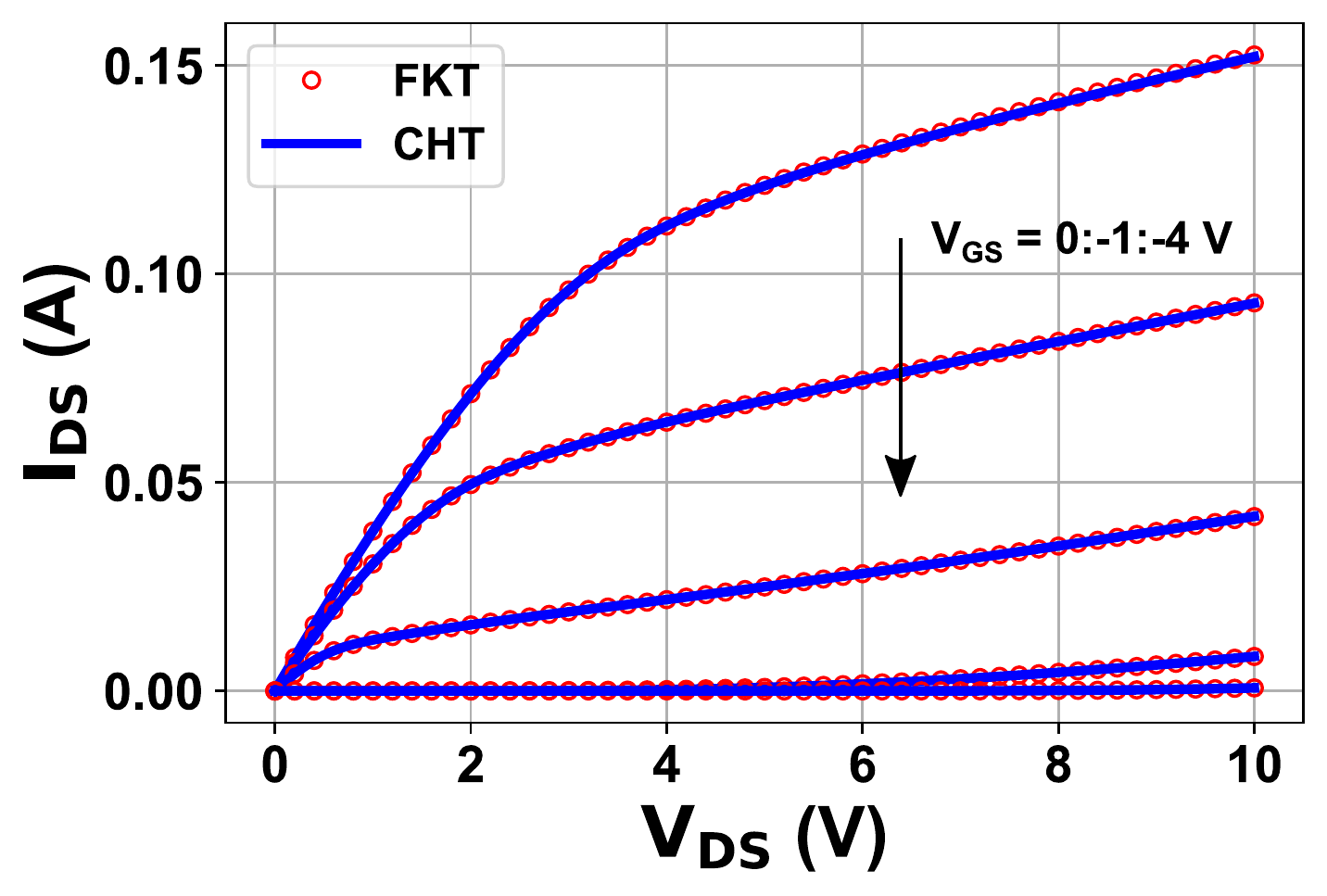}
    \caption{Drain current $I_D$ versus drain voltage $V_{DS}$ for different gate voltages $V_{GS}$ simulated by CHT and FKT with no electron heating, i.e.\ drift-diffusion.}
    \label{fig:idvd_DD}
\end{figure}


First, a drift-diffusion simulation was run using both FKT and CHT.
Once again, the material parameters used for the simulations and shown in Table~\ref{tab:material_pars} are kept identical. The voltage step size is fixed at 0.1 V for both the simulators. As seen in Fig. \ref{fig:idvd_DD}, both the simulators give nearly identical drain current. The minor differences in the obtained result can be attributed to the difference in the representation of the electron flux in CHT, (Eq. \ref{eq:Jn}), and the differences in the Scharfetter-Gummel discretization in the respective simulators.

\begin{figure}
    \centering
    \includegraphics[width=0.42\textwidth]{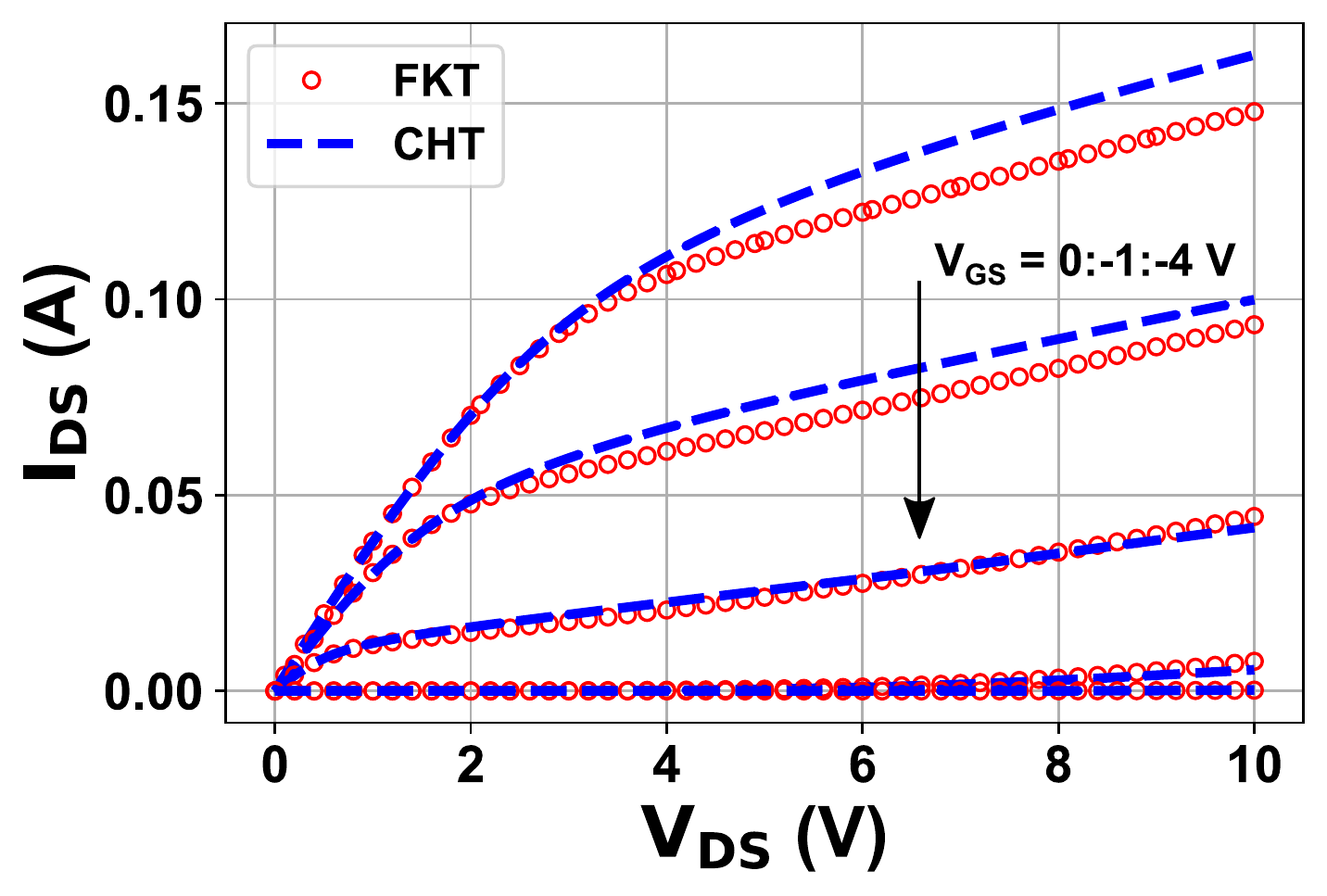}
    \caption{Drain current $\mathrm{I_D}$ versus drain voltage $\mathrm{V_{DS}}$ for different gate voltages $\mathrm{V_{GS}}$ simulated by CHT and FKT including hot electrons, $\tau_e=0.2$~ps.}
    \label{fig:idvd_Tn}
\end{figure}
Next, an energy transport simulation was run using the two simulators. The parabolic band approximation was used to model the dispersion relation with an electron effective mass of $0.2{m_0}$ ($m_0$ is the free electron mass). The energy relaxation time ($\tau_e$) is $0.2$ ps.~\cite{yeHotElectronRelaxation1999, zhangEnergyRelaxationHot2014} The CHT-specific hydrodynamic simulation parameters used are shown in Table~\ref{tab:sent_param}. For the FKT simulation, the voltage step size is kept fixed and unchanged at 0.1 V.  On the other hand, it was observed that an adaptive step size was required for the convergence of the CHT simulation. The maximum allowed voltage step size is set to 0.1~V, and the minimum allowed (initial) step size is lowered to 2.5~mV.

\begin{figure}
    \centering
    \includegraphics[width=0.45\textwidth]{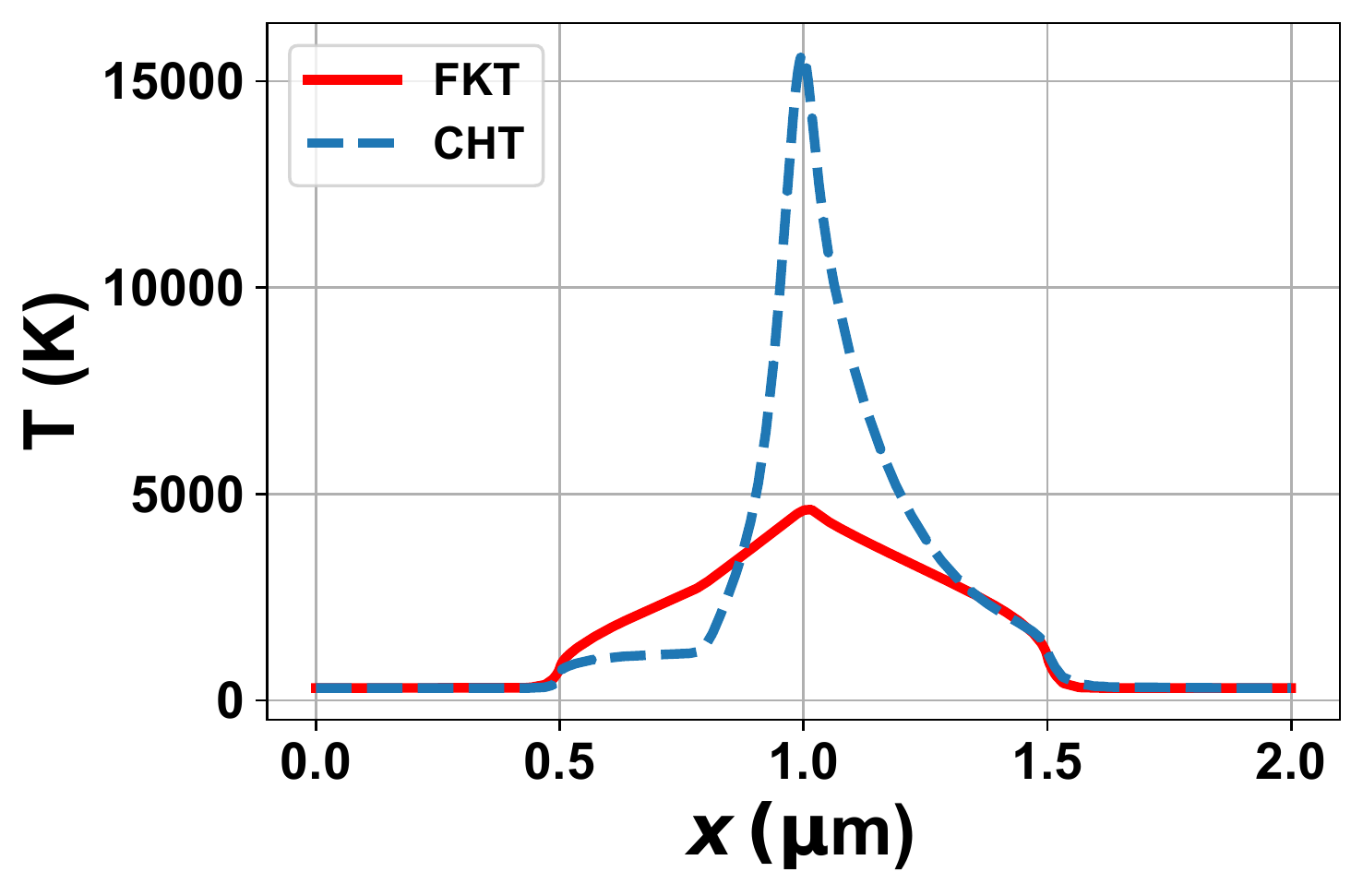}
    \caption{Comparison of electron temperatures produced by CHT and FKT along the x-direction (channel) at 6 nm below the AlGaN/GaN interface, for $V_{GS}=0$~V, $V_{DS}=10$~V\@, and $\tau_e = 0.2$ ps for both simulators.}
    \label{fig:Tn_x}
\end{figure}

The $I_D - V_{DS}$ characteristic and the electron temperature profile in the channel produced by the simulators is shown in Fig. \ref{fig:idvd_Tn} and Fig. \ref{fig:Tn_x}, respectively. It would seem unexpected that despite the higher electron temperature obtained by CHT, the drain current produced is also marginally higher than that produced by FKT. This can be explained by the fact that both the simulators are using a constant mobility. As a result, the scattering effects associated with high electron temperatures are not considered. Thus, in this case, higher electron temperatures signify higher carrier velocity with constant mobility, and consequently higher current.

As noted above, while the output characteristic obtained are quite similar, the electron temperature profile in the channel shows a marked difference between the two simulators. The electron temperatures produced by FKT are considerably lower than those from CHT, when $\tau_e = 0.2$ ps is chosen for both simulators. This is because of the significantly higher electronic heat flow into the ohmic contacts through the degenerately doped source/drain regions in FKT. More physically realistic simulations should include lattice heating, and FKT would transfer electron energy entering the ohmic mesh points directly into the semiconductor crystal, leading to lattice temperature hot spots at ohmic contact mesh points, particularly those on the drain contact. However, the simplified simulations presented here neglect lattice heating in order to highlight electron energy transport, and that electronic energy flowing into ohmic mesh points is removed instantly from the simulation domain.

\begin{figure*}
    \centering
    \includegraphics[width=0.8\textwidth]{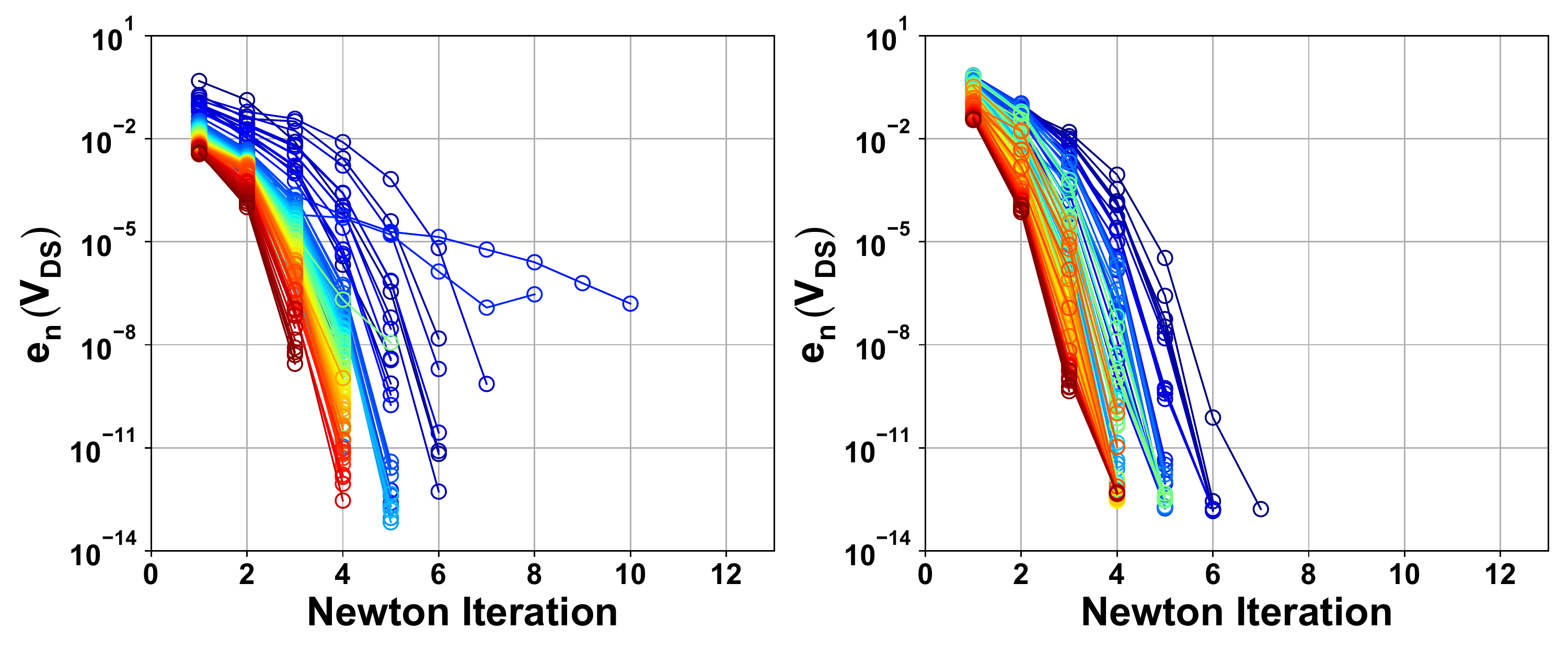}
    \caption{Calculated Newton solver residuals of CHT (left) and FKT (right) hot electron simulations with $\tau_e=0.2$~ps at $V_{GS}=-4$ V. The residuals are functions of the Newton iteration and the quiescent drain voltage.}
    \label{fig:residuals_Tn_0p2ps_Vgm4}
\end{figure*}
\begin{figure*}
    \centering
    \includegraphics[width=0.8\textwidth]{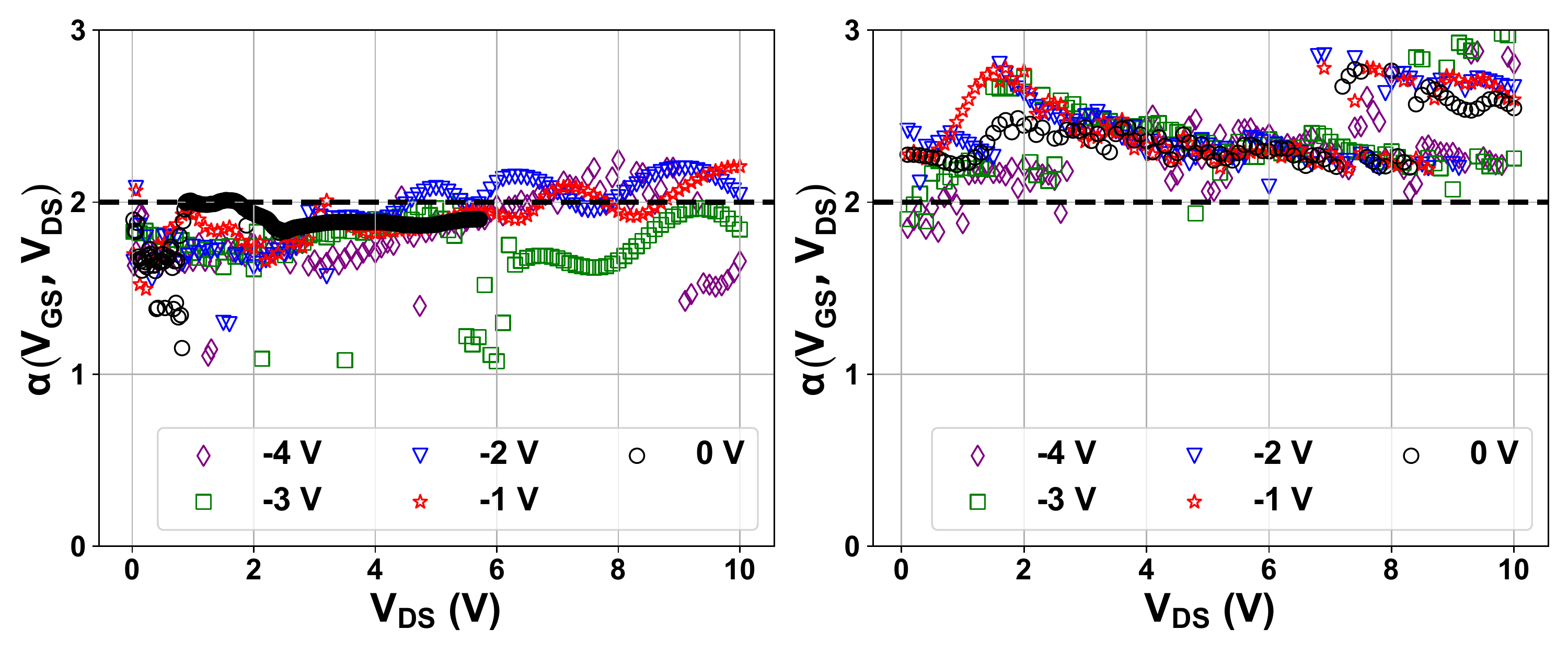}
    \caption{Estimated ROC of CHT (left) and FKT (right) hot electron simulations with $\tau_e=0.2$~ps and $V_{GS}$ ranging from -4\,V to 0\,V in 1\,V steps. The black dashed line indicates the region of ROC values representative of quadratic convergence.}
    \label{fig:roc_Tn_0p2ps_Vg_range}
\end{figure*}


\begin{figure}
    \centering
    \includegraphics[width=0.45\textwidth]{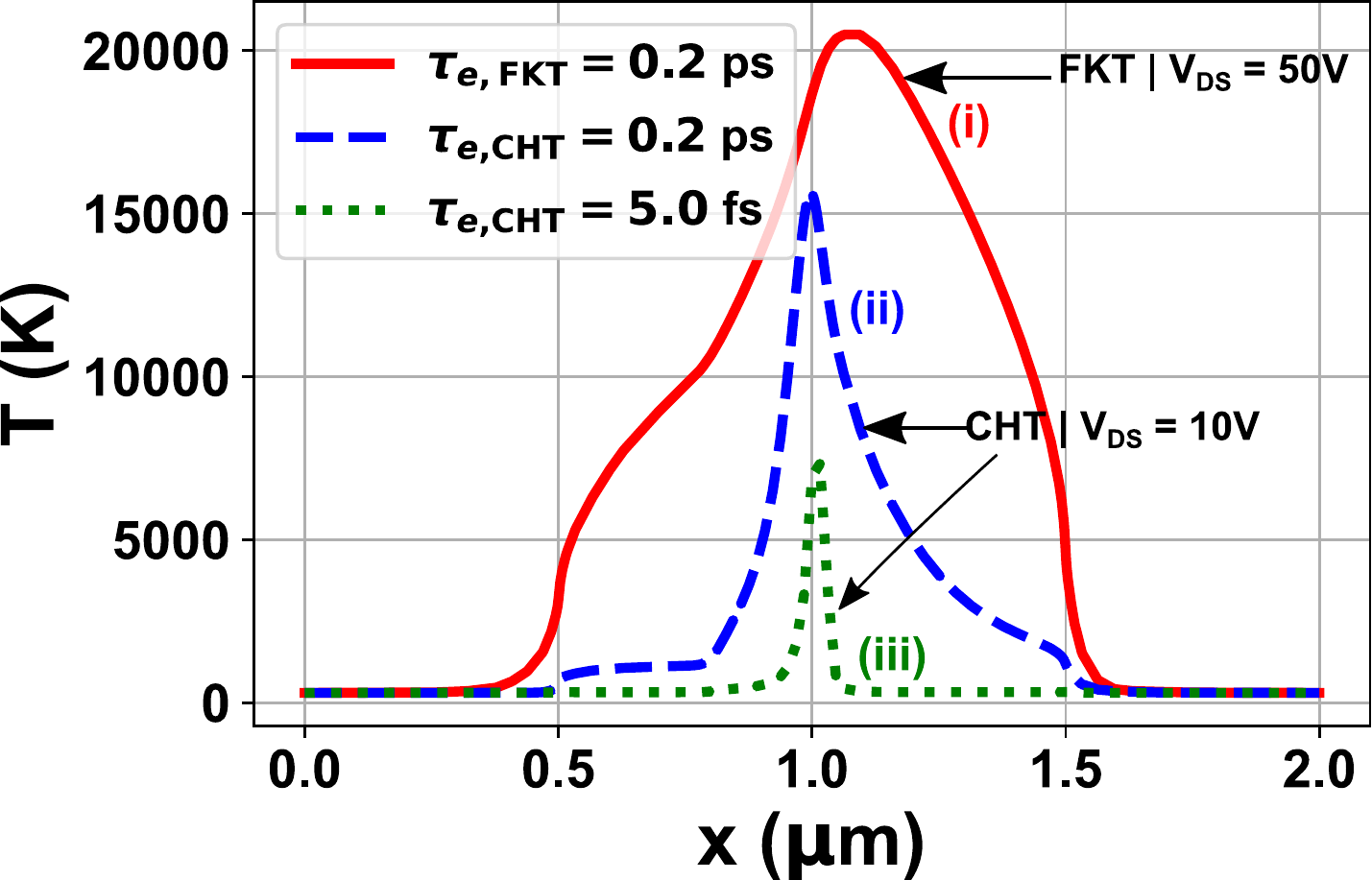}
    \caption{Comparison of electron temperatures produced by CHT and FKT along the x-direction (channel) at 6 nm below the AlGaN/GaN interface at $V_{GS}=0$\,V. (i) Temperature profile produced by FKT on increasing $V_{DS}$ to 50 V, at $\tau_e = 0.2$ ps. (ii) Temperature profile produced by CHT for $V_{DS} = 10$\, V and $\tau_e = 0.2$ ps. (iii) Temperature profile produced by CHT for $V_{DS} = 10$\, V and $\tau_e = 5.0$ fs.}
    \label{fig:comparable_Tn_x}
\end{figure}

We next analyze the convergence of the nonlinear solvers to explore other differences of the transport simulators. The outputs of the Newton solvers were stored to disk and post processed using custom Python codes. In particular, the Newton solver residuals were used to analyze the convergence of the transport simulators. These residuals are a function of Newton iteration and are computed at each quiescent gate and drain voltage. We define the Newton solver residual computed at the $n^{th}$ Newton iteration for a specific gate and drain quiescent bias as $e_n\left(V_{GS}, V_{DS}\right)$.

An example of Newton solver residuals is illustrated in Fig.~\ref{fig:residuals_Tn_0p2ps_Vgm4}. Here, residual values computed in the CHT device and FKT simulations are reported in the left and right of the figure, respectively. These residuals were computed at ${V_{GS}} = -4$\,V over the quiescent drain voltage sweep from ${V_{DS}} = 0$ to 10\,V. The FKT simulations exhibit excellent convergence characteristics for all quiescent drain voltage biases. However, as can be seen in Fig.~\ref{fig:residuals_Tn_0p2ps_Vgm4}, there are some instances of slow convergence exhibited by the CHT device simulator. For example, there is one Newton solver instance which takes 10 iterations to complete, and the final residual value is well above the typical value of $< 10^{-10}$.

A quantitative investigation of the convergence properties of the transport solvers is next presented. In particular, we analyze the rate of convergence (ROC), $\alpha$, of the numerical solvers. The ROC is calculated by \cite{senning2007}
\begin{equation}
\alpha\left(V_{GS}, V_{DS}\right) \approx \frac{\log \left|e_{n+1}\left(V_{GS}, V_{DS}\right)/e_n\left(V_{GS}, V_{DS}\right) \right|}{\log \left|e_{n}\left(V_{GS}, V_{DS}\right)/e_{n-1}\left(V_{GS}, V_{DS}\right) \right|}.
\end{equation}
The calculated ROCs processed from the CHT and FKT Newton solver residuals are illustrated in Fig. \ref{fig:roc_Tn_0p2ps_Vg_range}. These results indicate that FKT exhibits at least quadratic convergence ($\alpha = 2$) for all the drain biases. It should be noted that there exists some FKT ROC values which fall slightly below $\alpha = 2$. These values, on the order of $\alpha = 1.8$, are the result of numerical precision of the solver. For example, the set of Newton solver residuals which contain a value at the 7$^{th}$ iteration shed light on the impact of numerical precision of the solver. The slope of the residual curve changes between iteration 6 and 7 due to hitting the numerical noise floor in the solver. This change in slope affects the estimated ROC value and would not occur if there was not a numerical noise floor in the calculations. The ROC of CHT is also quadratic at several quiescent bias solutions. However, the ROC of the CHT simulation is generally super-linear. 


As noted above, the two simulators exhibit different convergence characteristics, with FKT  generally displaying quadratic convergence, compared to CHT showing super-linear convergence. In addition, the simulators also produced widely different electron temperatures in the channel. It would be interesting to determine if this difference in the convergence behavior can be attributed to the much higher electron temperatures obtained from the CHT simulations. This can be verified by running the simulations to control for the temperature difference. Therefore, we analyze the convergence characteristics by changing the simulation setup so as to obtain similar peak electron temperatures. 

We investigate two methods to obtain similar channel electron temperatures from the two simulators. First, we try to reduce the electron temperatures produced by CHT, to the approximated range of temperatures yielded by the unaltered FKT simulation. For that purpose, we reduce the $\tau_e$ used in CHT to an artificially low value. Reducing the energy relaxation time allows the electrons to rapidly lose their energy to the lattice, thereby reducing the electron temperature. For the second method, we try to increase the drain bias in the FKT simulation, keeping the CHT setup unchanged. The exceedingly high fields produced should result in hot electrons with temperatures comparable to CHT.

\begin{figure*}
    \centering
    \includegraphics[width=0.8\textwidth]{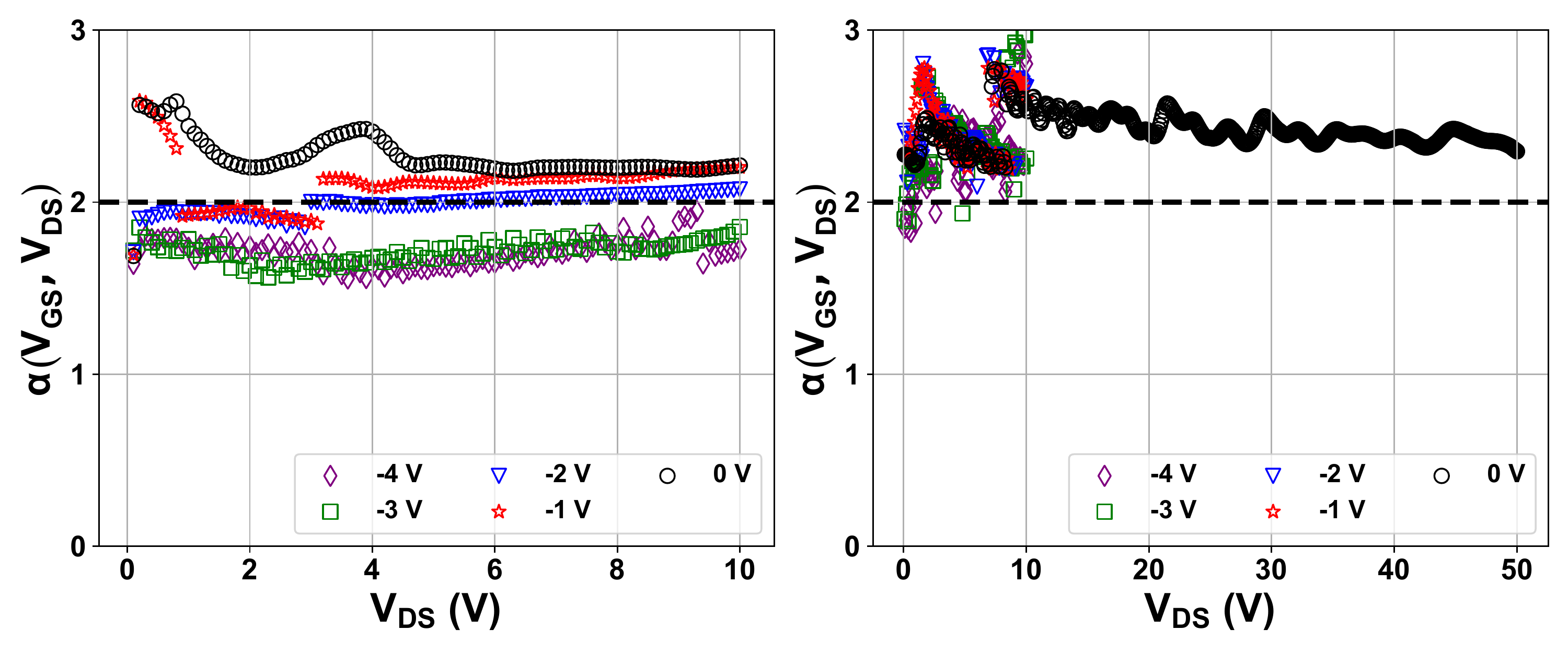}
    \caption{Estimated ROC of CHT (left) hot electron simulations with $\tau_e=0.005$~ps and $\sigma_{PZ} = 9 \times 10^{12} \mathrm{cm}^{-2}$. Estimated ROC of FKT (right) hot electron simulations with $\tau_e=0.2$~ps, $\sigma_{PZ} = 9 \times 10^{12} \mathrm{cm}^{-2}$ and drain voltage up to 50\,V. The symbols correspond to $V_{GS}$, as indicated in the legends.}
    \label{fig:roc_Tn_5fs_high_vds}
\end{figure*}

Accordingly, we first perform the CHT simulation by reducing the $\tau_e$ to $5.0$ fs, while keeping the $\tau_e$ in FKT unchanged at $0.2$ ps. Fig.~\ref{fig:comparable_Tn_x} shows that the peak electron temperature in the channel is in the same range of magnitude as FKT. The convergence characteristic for the CHT simulation is shown in Fig. \ref{fig:roc_Tn_5fs_high_vds} (left). We see that the ROC is generally super-linear, and quadratic or above for some of the gate biases. Therefore, reducing the electron temperature did not significantly change the CHT convergence rate, and it still shows a slower ROC compared to FKT.

In addition to simulations that lowered electron temperature in CHT to approximate the original FKT temperatures in Fig.~\ref{fig:Tn_x}, further FKT simulations also increased electron temperatures to approximate the original CHT temperatures in Fig.~\ref{fig:Tn_x}. Because $T_n$ was reduced in CHT simulations by decreasing $\tau_e$ far below 0.2~ps, it might seem natural to increase $\tau_e$ far above 0.2~ps in FKT to produce much higher $T_n$. However, when $\tau_e$ increases excessively, the electronic heat flow to the ohmic contacts in FKT limits further increase in channel electron temperature. A more effective way to heat FKT electrons to temperatures comparable to the CHT temperatures in Fig.~\ref{fig:Tn_x} is to simply drive the FKT simulations to higher drain voltages. Thus, the FKT simulation was run up to ${V_{DS}} = 50$ V at ${V_{GS}} = 0$\,V. Fig.~\ref{fig:comparable_Tn_x} shows that the peak electron temperature obtained for this setup is comparable to that obtained from the original CHT simulation. We observe that FKT continues to show excellent quadratic convergence, as seen in Fig.~\ref{fig:roc_Tn_5fs_high_vds} (right). Despite increasing the electron temperature by applying high fields, FKT continues to show quadratic convergence compared to the super-linear convergence of CHT.

Finally, changing the polarization sheet charge density ($\sigma_{\mathrm{PZ}}$) at AlGaN/GaN interface, changes the field at the interface. The large electric fields that arise place further burden on the solvers and affect the convergence behavior. Therefore, a series of simulations were run by increasing $\sigma_{\mathrm{PZ}}$, with $\tau_e = 0.2$ ps for both the simulators, to characterize the convergence behavior. As summarized in TABLE \ref{tab:roc_summary}, FKT displays greater than quadratic ROC on average for all values of $\sigma_{\mathrm{PZ}}$, while CHT shows a super-linear ROC that decreases as $\sigma_{\mathrm{PZ}}$ is increased. Furthermore, the CHT simulations failed to converge for $\sigma_{\mathrm{PZ}} \geq 15 \times 10^{12}$~$\mathrm{cm^{-2}}$, which is indicated as "N/A" in TABLE \ref{tab:roc_summary}. For the simulations to converge, we observed that $\tau_e$ had to be reduced to an unphysically low value of $5.0$~fs.

\begin{table}[]
\caption{\label{tab:roc_summary} Summary of CHT device and FKT ROC values. "N/A" refers to simulations that failed to converge.}
    \begin{ruledtabular}
    \begin{tabular}{c  c  c }
        \multirow{2}{*}{ $\sigma_{PZ} \left(\times 10^{12} \hspace{0.1cm} \mathrm{cm}^{-2}\right)$} & 
        \multicolumn{2}{ c }{mean$\{\alpha\left(V_{GS}, V_{DS}\right)\}$}\\ 
        & CHT & FKT \\
        \hline
        9 & 1.856 & 2.399 \\ 
        10 & 1.809 & 2.387 \\ 
        11 & 1.044 & 2.392 \\ 
        12 & 1.484 & 2.409 \\ 
        13 & 1.510 & 2.403 \\ 
        14 & 1.560 & 2.397 \\ 
        15 & N/A & 2.391 \\ 
        16 & N/A & 2.387 \\ 
        17 & N/A & 2.384 \\ 
        18 & N/A & 2.388 \\ 
    \end{tabular}
    \end{ruledtabular}
\end{table}

\subsection{Time-Dependent Simulations}
\begin{figure*}
\centering
\includegraphics[width=0.42\textwidth]{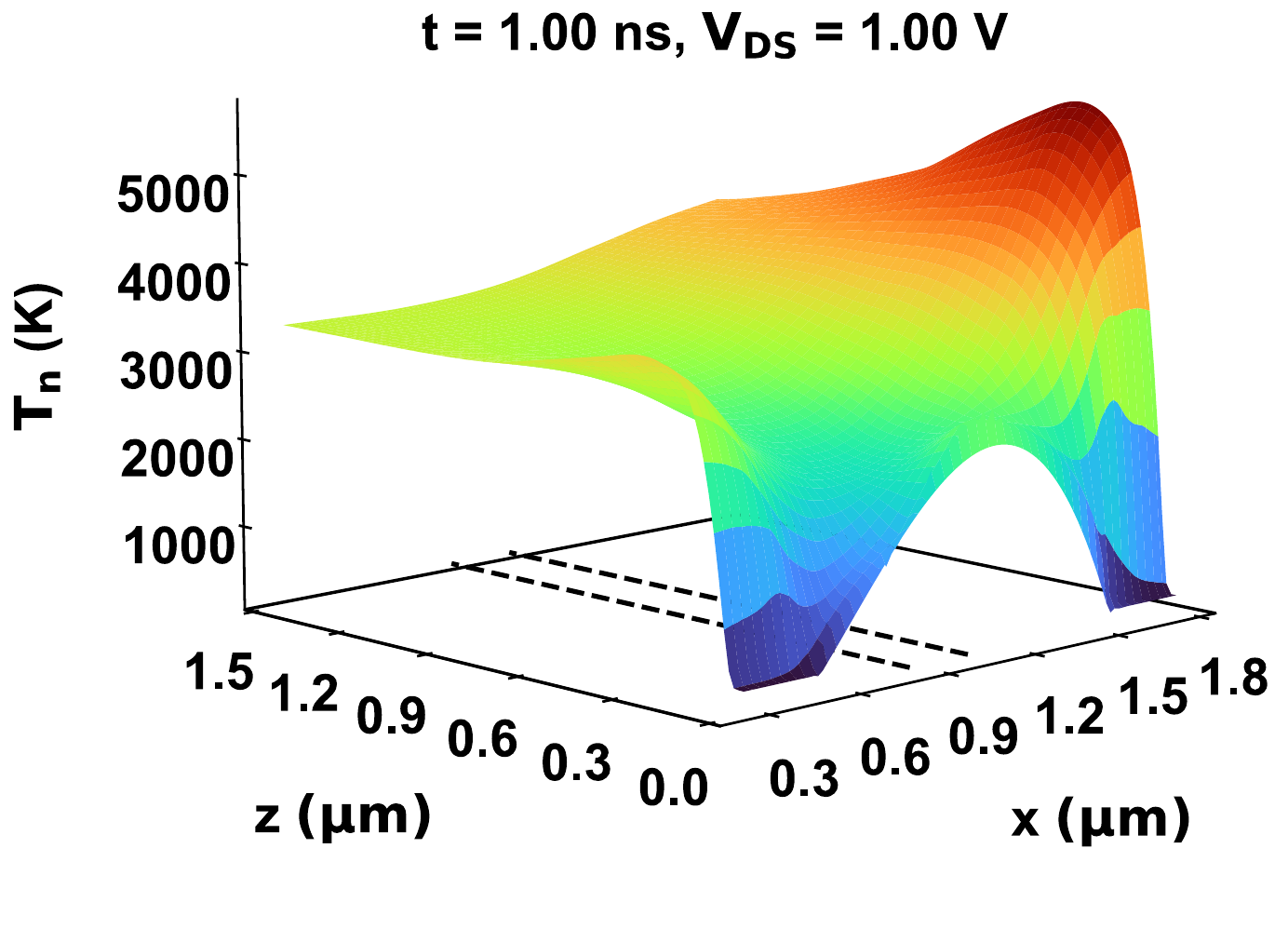}
\includegraphics[width=0.42\textwidth]{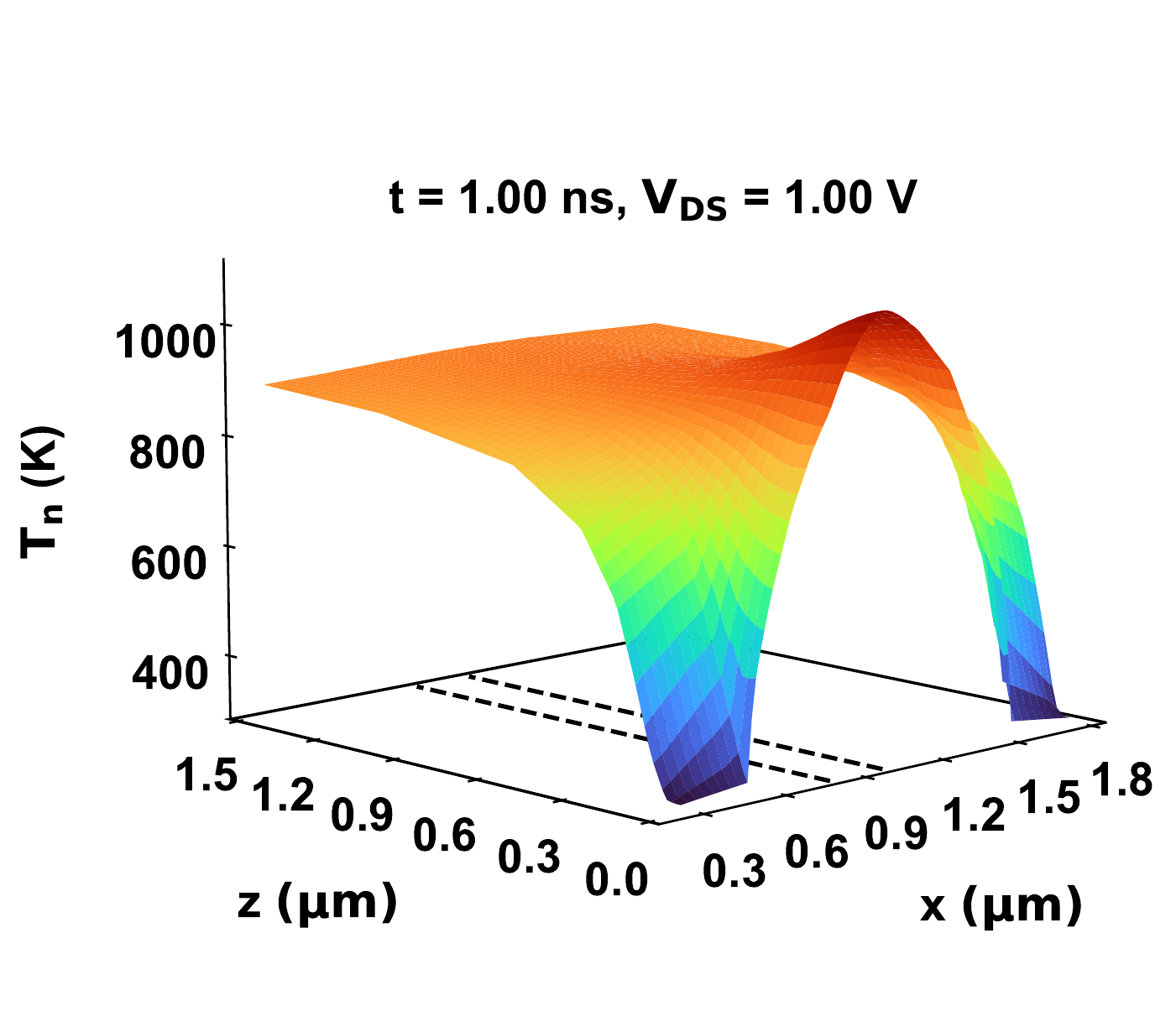}
\caption{Electron temperatures $T_n$ across the transistor for
$V_{GS}=0$ when $V_{DS}$ is increased from 0 to 1~V in 1~ns as
computed with the CHT model (left) and the FKT model (right).
The AlGaN barrier/GaN channel interface is located at z = 0, and
the position of the gate in the x-direction is represented by the dotted lines. (Multimedia view)
}
\label{fig:xsTn100}
\end{figure*}

Because the above results suggest the different CHT and FKT models may
have led to different numerical convergence behavior,
transient simulations were performed to test for additional
differences in behavior. Starting from the thermal equilibrium initial
condition, the drain voltage $V_{DS}$ was ramped from 0 to 1~V in
1~ns, and the loss of electron energy to the semiconductor crystal
lattice was set to zero in order to highlight differences in
real-space energy transport. The electron temperatures across the
device computed by CHT
and FKT are shown in animated Figs.~\ref{fig:xsTn100} (Multimedia view) (left) and
(right), respectively.
\begin{figure*}
\centering
\includegraphics[width=0.42\textwidth]{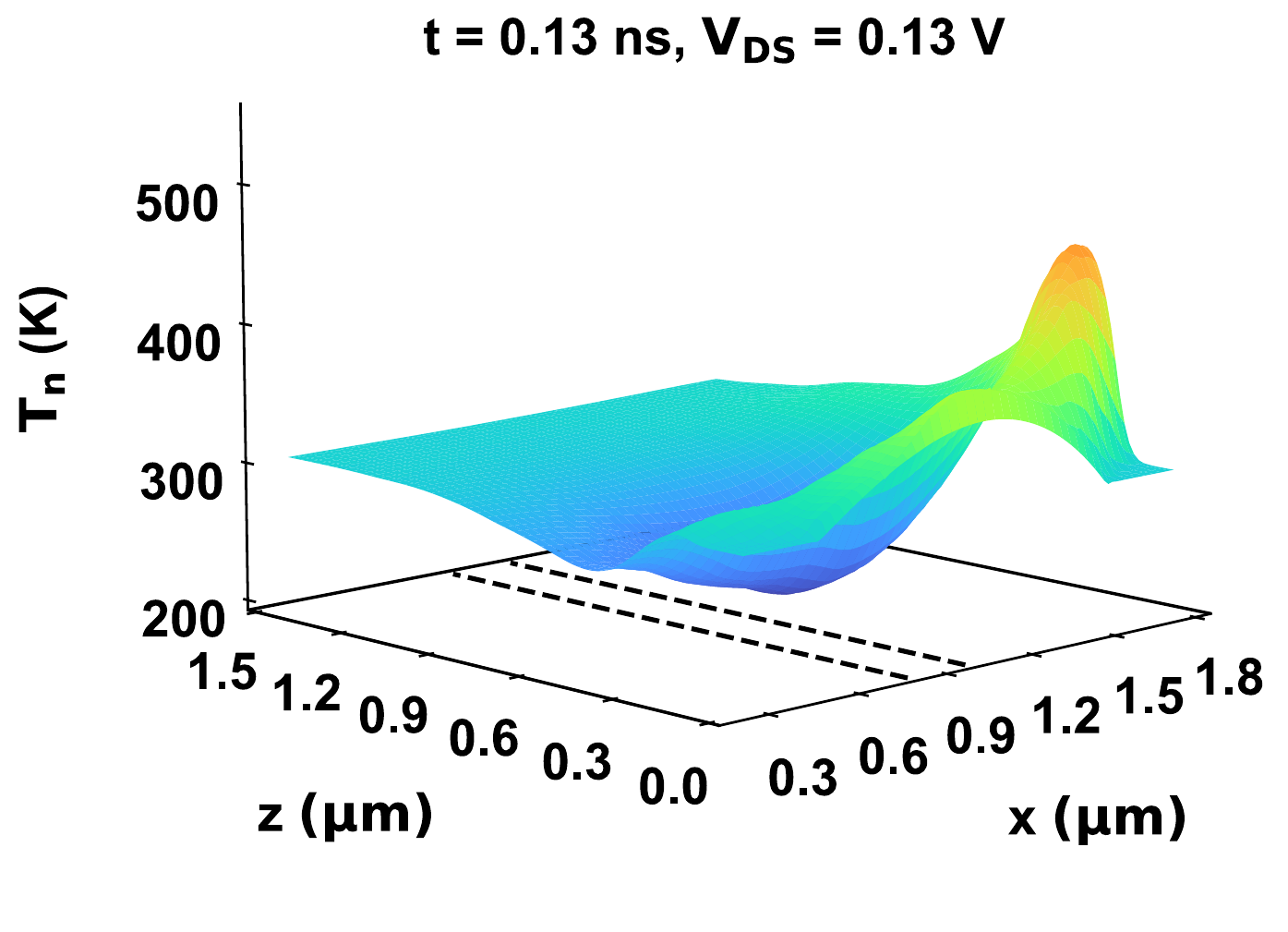}
\includegraphics[width=0.42\textwidth]{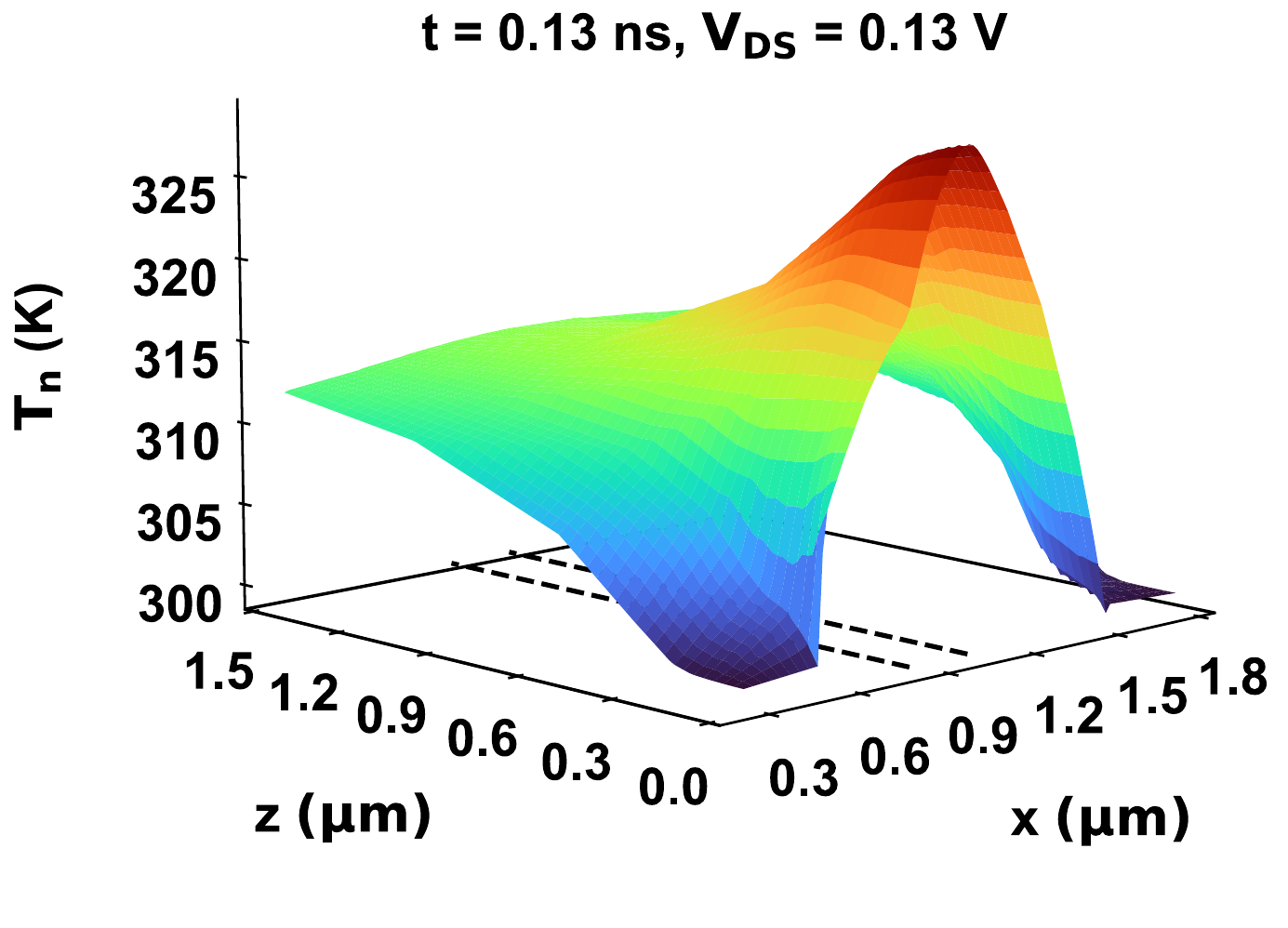}
\caption{Electron temperatures $T_n$ across the transistor computed by
with the CHT model (left) and the FKT model (right) over the time span
$0<t<0.15$~ns as the drain voltage $V_{DS}$ changed at a rate of
1~V/ns.
The AlGaN barrier/GaN channel interface is located at z = 0, and
the position of the gate in the x-direction is represented by the dotted lines. (Multimedia view)
}
\label{fig:xsTn015}
\end{figure*}
As expected from the previous static results, the
electrons heat up to higher temperatures in the CHT
calculation. However, the peak CHT temperatures are
relatively distant from the gate, appearing underneath the drain
contact. The peak FKT temperatures, on the other hand, appear at the
corner of the gate closer to the drain. Because most of the $V_{DS}$
voltage drop occurs at this corner of the gate, it is unclear why the
peak CHT electron temperature would not also occur in this location.
In addition to locations of peak electron temperature, the transient
simulations also produced other differences that become apparent when
focusing on the shorter time scale $0<t<0.15$~ns. The animated
Figs.~\ref{fig:xsTn015} (Multimedia view) (left) and (right)
show these dynamic electron
temperatures across the device as produced by the CHT and FKT models,
respectively. As in the case of the longer time scales, FKT shows
electron temperatures increasing monotonically in time with the peak
located at the drain-side corner of the gate. The CHT model again
produces a peak electron temperature occurring away from the gate and
underneath the ohmic drain contact, but it also shows electron
temperatures in the GaN buffer region under the gate decreasing
significantly below the 300~K ambient temperature. As these electron
dynamics occur at constant volume in the absence of physical work, it
is not immediately clear what physical mechanism could cause electron
temperatures inside the transistor to fall below the ambient
temperature. The fundamentally different physical behaviors revealed by these dynamically evolving transient simulations may help explain the different numerical convergence characteristics previously observed for the CHT and FKT static simulations.

\section{Conclusion}
Deterministic Boltzmann solvers make use of the moments of the BTE to model the charge transport in semiconductors. Hydrodynamic model and FKT are both deterministic solvers that belong to this class of energy transport models. They differ in the way they treat the electronic heat flow. While hydrodynamic models utilize an electron thermal conductivity, FKT employs the heat capacity of an ideal Fermi gas.

In this work, we compared the hydrodynamic model implemented in CHT with the FKT model. A GaN HEMT structure was used to demonstrate the simulators. It was observed that both the simulators gave reasonably similar non-equilibrium output characteristics. However, the electron temperatures in the channel, close to the AlGaN/GaN interface, were widely different. The rate of convergence of the solvers were also different, with FKT showing quadratic convergence or higher, and CHT showing super-linear convergence.

Next, we investigated if the difference in the convergence characteristics can be attributed to the different electron temperatures obtained. We employed the two methods to ensure that the two simulators produce similar temperatures. First, we reduced the energy relaxation time used in CHT simulation, keeping the FKT setup unchanged. Second, we increased the drain bias for the FKT simulation, producing hot electrons with temperature comparable to the unchanged CHT simulation. We then compared the convergence behavior of FKT and CHT in each of the above mentioned cases. As before, FKT continued to show greater than quadratic rate of convergence, and CHT showed super-linear convergence. We can conclude that the different convergence behavior is due to the difference in the transport models and the description of the electronic heat flow, rather than the difference in the electron temperatures produced. Additionally, simulations were run after changing the polarization sheet charge density at the AlGaN/GaN interface. The results show that FKT
exhibited rapid quadratic convergence for all values chosen, while CHT failed to converge for large values of the polarization charge.

To further explore the possible underlying causes of different numerical convergence, additional large signal transient simulations were performed, and the evolving electron temperatures inside the device, as produced by the CHT and FKT models, were examined. During the transients, FKT electron temperatures throughout the HEMT structure increased monotonically with drain voltage, and the peak temperature consistently appeared at the drain side of the gate where the lateral electric fields are largest and most of the drain voltage drop occurs. CHT, on the other hand, produced peak electron temperatures located away from the gate and in the degenerately doped drain region. Also, CHT electron temperatures did not increase monotonically throughout the device during the transient drain volt sweep. At early stages of the voltage sweep, CHT produced electron temperatures in the buffer region underneath the gate that dropped nearly 100 K below the ambient temperature. Because these electron dynamics occurred under constant volume conditions, there is no apparent physical mechanism that may have caused this electron cooling. Instead, it may reveal some fundamental differences between the CHT and FKT models, particularly their different treatments of electronic heat flow. These results also suggest that FKT could potentially provide a robust and effective means for simulating GaN HEMT structures for certain applications including RF power electronics.


\begin{acknowledgments}

This work was supported by AFOSR Grant No.\ LRIR 21RYCOR073.

\end{acknowledgments}

\section*{Data Availability Statement}
The data that support the findings of this study are available from the corresponding author upon reasonable request.
\bibliography{biblio1, AT_refs, Nick_refs}

\end{document}